\documentstyle[12pt]{article}

\catcode`\@=11
\long\def\@makefntext#1{
\protect\noindent \hbox to 3.2pt {\hskip-.9pt  
$^{{\ninerm\@thefnmark}}$\hfil}#1\hfill}                

\def\@makefnmark{\hbox to 0pt{$^{\@thefnmark}$\hss}}  
        
\def\ps@myheadings{\let\@mkboth\@gobbletwo
\def\@oddhead{\hbox{}
\rightmark\hfil\ninerm\thepage}   
\def\@oddfoot{}\def\@evenhead{\ninerm\thepage\hfil
\leftmark\hbox{}}\def\@evenfoot{}
\def\sectionmark##1{}\def\subsectionmark##1{}}

\renewcommand{\thefootnote}{\fnsymbol{footnote}}

\newcounter{sectionc}\newcounter{subsectionc}\newcounter{subsubsectionc}
\renewcommand{\section}[1] {\vspace*{0.6cm}\addtocounter{sectionc}{1} 
\setcounter{subsectionc}{0}\setcounter{subsubsectionc}{0}\noindent 
        {\normalsize\bf\thesectionc. #1}\par\vspace*{0.4cm}}
\renewcommand{\subsection}[1] {\vspace*{0.6cm}\addtocounter{subsectionc}{1} 
        \setcounter{subsubsectionc}{0}\noindent 
        {\normalsize\it\thesectionc.\thesubsectionc. #1}\par\vspace*{0.4cm}}
\renewcommand{\subsubsection}[1] {\vspace*{0.6cm}\addtocounter{subsubsectionc}{1}
        \noindent
{\normalsize\rm\thesectionc.\thesubsectionc.\thesubsubsectionc. 
        #1}\par\vspace*{0.4cm}}

\newcounter{appendixc}
\newcounter{subappendixc}[appendixc]
\newcounter{subsubappendixc}[subappendixc]

\renewcommand{\appendix}[1] {\vspace*{0.6cm}
        \refstepcounter{appendixc}
        \setcounter{figure}{0}
        \setcounter{table}{0}
        \setcounter{equation}{0}
        \renewcommand{\thefigure}{\Alph{appendixc}.\arabic{figure}}
        \renewcommand{\thetable}{\Alph{appendixc}.\arabic{table}}
        \renewcommand{\theappendixc}{\Alph{appendixc}}
        \renewcommand{\theequation}{\Alph{appendixc}.\arabic{equation}}
        \noindent{\bf Appendix \theappendixc #1}\par\vspace*{0.4cm}}

\def\abstracts#1{{
       
\centering{\begin{minipage}{12.2truecm}\footnotesize\baselineskip=12pt\noindent
        \centerline{\footnotesize ABSTRACT}\vspace*{0.3cm}
        \parindent=0pt #1
        \end{minipage}}\par}} 

\newcounter{itemlistc}
\newcounter{romanlistc}
\newcounter{alphlistc}
\newcounter{arabiclistc}

\newcommand{\fcaption}[1]{
        \refstepcounter{figure}
        \setbox\@tempboxa = \hbox{\footnotesize Fig.~\thefigure. #1}
        \ifdim \wd\@tempboxa > 6in
           {\begin{center}
        \parbox{6in}{\footnotesize\baselineskip=12pt Fig.~\thefigure. #1}
            \end{center}}
        \else
             {\begin{center}
             {\footnotesize Fig.~\thefigure. #1}
              \end{center}}
        \fi}

\newcommand{\tcaption}[1]{
        \refstepcounter{table}
        \setbox\@tempboxa = \hbox{\footnotesize Table~\thetable. #1}
        \ifdim \wd\@tempboxa > 6in
           {\begin{center}
        \parbox{6in}{\footnotesize\baselineskip=12pt Table~\thetable. #1}
            \end{center}}
        \else
             {\begin{center}
             {\footnotesize Table~\thetable. #1}
              \end{center}}
        \fi}

\def\@citex[#1]#2{\if@filesw\immediate\write\@auxout
        {\string\citation{#2}}\fi
\def\@citea{}\@cite{\@for\@citeb:=#2\do
        {\@citea\def\@citea{,}\@ifundefined
        {b@\@citeb}{{\bf ?}\@warning
        {Citation `\@citeb' on page \thepage \space undefined}}
        {\csname b@\@citeb\endcsname}}}{#1}}

\newif\if@cghi
\def\cite{\@cghitrue\@ifnextchar [{\@tempswatrue
        \@citex}{\@tempswafalse\@citex[]}}
\def\citelow{\@cghifalse\@ifnextchar [{\@tempswatrue
        \@citex}{\@tempswafalse\@citex[]}}
\def\@cite#1#2{{$\null^{#1}$\if@tempswa\typeout
        {IJCGA warning: optional citation argument 
        ignored: `#2'} \fi}}

 1
 1
 1

\font\ninerm=cmr9

\def\lsim{\mathrel{\mathop{\kern 0pt <}\limits_{\displaystyle\sim}}}
\def\gsim{\mathrel{\mathop{\kern 0pt >}\limits_{\displaystyle\sim}}}
\begin{document}

\centerline{\normalsize\bf POWER CORRECTIONS AND LANDAU SINGULARITY
\footnote
{Research supported in part by the EC program `Human Capital and Mobility',
Network `Physics at High Energy Colliders', contract CHRX-CT93-0357 (DG12 COMA).}}
\baselineskip=22pt

\vspace{0.6cm}
\centerline{\footnotesize Georges GRUNBERG}

\baselineskip=13pt
\centerline{\footnotesize\it Centre de Physique Th\'eorique de l'Ecole
 Polytechnique\footnote{CNRS UPRA 0014}}
\baselineskip=12pt
\centerline{\footnotesize\it 91128 Palaiseau Cedex - France}
\vspace{0.6cm}
\centerline{\footnotesize E-mail: grunberg@pth.polytechnique.fr}

\vspace{0.9cm}
\abstracts{\normalsize 
In the dispersive approach of Dokshitzer, Marchesini and Webber,  standard
power-behaved contributions of infrared origin are described with the notion of an infrared
regular QCD  coupling. I argue that their framework suggests the existence of
non-standard    contributions, arising from short distances (hence unrelated to renormalons
and the operator product expansion),  which  appear in the process of removing the Landau
singularity of the perturbative coupling. A natural definition of an infrared finite perturbative
coupling is suggested within the dispersive method. Implications for the tau hadronic width
and the lattice determination of the gluon condensate, where ${\cal O}(1/Q^2)$ contributions can
be generated, are pointed out.} 
\vspace{5cm}
CPTh/S 505.0597
\hspace{9cm}
May 1997

\newpage 
\pagestyle{plain}
\normalsize\baselineskip=15pt
\setcounter{footnote}{0}
\renewcommand{\thefootnote}{\alph{footnote}}
\section{Introduction} The study of power corrections in QCD has been the subject of active
investigations in recent years. Their importance for a precise determination of $\alpha_s$ has
been recognized, and  various  techniques (renormalons [1], finite gluon mass [2-6], dispersive
approach [7]  ) have been devised to cope with situations where the standard operator product
expansion (OPE) does not apply.  Standard power-behaved contributions in QCD arise from
non-perturbative effects at low scale, reflecting the non-trivial vacuum structure. In this
paper, I  concentrate on the dispersive approach [7], based on the notion of an infrared (IR)
regular [8] (see also [9])  QCD coupling, where  a non-perturbative contribution to the coupling,
essentially restricted to low scales, parametrizes the  power corrections. I point out that
within this framework,  it is  very natural to expect the existence of new type of power
contributions {\em of ultraviolet (UV) origin, hence not controlled by the OPE}, related to the
removal of the IR Landau singularity presumably present in the perturbative part of the coupling.
After a brief review of the dispersive approach (section 2), a simple ``dispersive'' method
[10-13] (see also [14]) of removing the Landau singularity is suggested in section 3 as a
convenient definition of a `` regularized'' perturbative coupling; the full IR regular  coupling
is then obtained as the sum of the `` regularized perturbative coupling'' and  of the ``genuine''
non-perturbative piece  . The  former differs from the perturbative coupling by power
corrections  which are computed in  section 4 and 5 using Borel transform techniques.  In sharp
contrast to the genuine non perturbative part of the coupling, these corrections are not
restricted to  low energy , and can thus induce ``perturbative'' power contributions of
ultraviolet origin  in Euclidean observables,  considered in section 6.1 .  The
``non-perturbative'' part of the power corrections, induced by the corresponding piece of the 
coupling, is  discussed in section 6.2, and the framework of [7] is extended in a straightforward
way by  relaxing the assumption that  this piece is confined to the IR region. Section 7 deals
with Minkowskian observables . As a sample of applications, I discuss briefly in section 8
inclusive
$\tau$-decay, and the gluon condensate on the lattice. A critical assessment and concluding
remarks are given in section 9. Some more technical issues are developped in two appendices. In
particular, in Appendix A the stability [12] against higher order corrections of the value of the
IR fixed point of the `` dispersively regularized'' perturbative coupling  is proved within some
restrictions.

\section{Dispersive approach to power corrections}

Consider the contribution to an Euclidean (quark dominated) observable arising from dressed
single gluon virtual exchange, which takes the generic form (after subtraction of the Born term):
$$D(Q^2) =\int_{0}^\infty{dk^2\over k^2}\ \alpha_s(k^2)\ \varphi\left(k^2\over
Q^2\right)\eqno(2.1)$$ The IR regularity of the coupling is implemented through a dispersion
relation:
\begin{eqnarray*}\ \ \ \ \ \ \ \ \ \ \alpha_s(k^2) &=& -\int_0^\infty{d\mu^2\over\mu^2+k^2}\
\rho(\mu^2)\ \ \ \ \ \ \ \ \ \ \ \ \
\ \ \ \ \ \ \ \ \ \ \ \ \ \ \ \ \ \ \ \ \ \ \ \ \ \ \ \ \ \ \ \ \ \ \ \ \ \ \ (2.2a)\\
&\equiv&k^2\int_0^\infty{d\mu^2\over(\mu^2+k^2)^2}\ \alpha_{eff}(\mu^2)\ \ \ \ \ \ \ \ \ \ \ \ \
\ \ \ \ \ \ \ \ \ \ \ \ \ \ \ \ \ \ \ \ \ \ \ \ \ \ \ \ \ \ \ (2.2b)
\end{eqnarray*} where $\rho(\mu^2) = -\frac{1} {2\pi i} Disc\{\alpha_s(-\mu^2)\}\equiv -\frac{1}
{2\pi i}\{\alpha_s\left[-(\mu^2+i\epsilon)\right]-\alpha_s\left[-(\mu^2-i\epsilon)\right]\}$ is
the time like ``spectral density'', and the``effective coupling'' 
$\alpha_{eff}(\mu^2)$   is defined  by:
$$ {d\alpha_{eff}\over d\ln \mu^2}=\rho(\mu^2) \eqno(2.3)$$ i.e.
$$\alpha_{eff}(\mu^2)=-\int_{\mu^2}^\infty{d\mu'^2\over\mu'^2}\
\rho(\mu'^2)\eqno(2.4)$$ For small $\alpha_s$, $\alpha_{eff}(\mu^2)=\alpha_s(\mu^2)+{\cal
O}(\alpha_s^3)$. Eq.(2.2) guarantees the absence of Landau singularity in the whole first sheet
of the complex $k^2$ plane. The coupling $\alpha_s(k^2)$ might be understood as a universal
``physical'' QCD coupling (not to be confused with e.g. the MS coupling), an analogue of the
Gell-Mann - Low QED effective charge , hopefully defined through an extension to QCD of the QED
`` dressed skeleton expansion'' [15,16]: such a program, which would give a firm field
theoretical basis to the ``naive non-abelization'' procedure [17,6,18] familiar in renormalons
calculations, has been initiated in [19] (a different ansatz is however suggested in [7]). In the
``large $\beta_0$'' limit of QCD, as implemented through the ``naive non-abelization'' procedure
, $\alpha_s(k^2)$ then coincides with the V-scheme coupling [15] ( but differs [19] from it at
finite
$\beta_0$). 

\noindent Inserting eq.(2.2) into eq.(2.1) one gets:
\begin{eqnarray*}\ \ \ \ \ \ \ \ \ \ \ \ \ \ D(Q^2) &=& \int_0^\infty{d\mu^2\over\mu^2}\
\rho(\mu^2)\ \left[{\cal F}({\mu^2\over Q^2})-{\cal F}(0)\right] \ \ \ \ \ \ \ \ \ \ \ \ \ \ \ \
\ \ \ \ \ \ \ \ \
\ \ \ \ \ (2.5a)\\ &\equiv&\int_0^\infty{d\mu^2\over\mu^2}\ \alpha_{eff}(\mu^2)\ \dot{{\cal
F}}\left({\mu^2\over Q^2}\right)\ \ \ \ \ \ \ \ \ \ \ \ \ \ \ \ \ \ \ \ \ \ \ \ \ \ \ \ \ \
\ \ \ \ \ \ \ \ (2.5b)\end{eqnarray*}  where ${\cal F}$ is the ``characteristic function'' [7]:
$${\cal F}({\mu^2\over Q^2}) = \int_{0}^\infty{dk^2\over k^2+\mu^2}\ \varphi\left(k^2\over
Q^2\right)\eqno(2.6)$$ and $\dot{{\cal F}}\equiv - d{\cal F}/d\ln \mu^2$. Eq.(2.6) shows that the
``characteristic  function'' is just the ${\cal O}(\alpha_s)$ Feynman diagram computed with a
finite gluon mass $\mu^2$ [4,6], and that the Feynman diagram kernel 
$\varphi\left(k^2\over Q^2\right)$ (also called ``distribution function'' in [18]) is
proportionnal to the time-like discontinuity of ${\cal F}$. 

The authors of [7] moreover suggest that $\alpha_s(k^2)$ comprises both a ``perturbative'' and a
``non-perturbative'' part: 
$$\alpha_s(k^2)=``\alpha_s^{PT}(k^2)"+\delta\alpha_s^{NP}(k^2)
\eqno(2.7)$$ and similarly:
$$\alpha_{eff}(\mu^2)  = ``\alpha_{eff}^{PT}(\mu^2)"+\delta\alpha_{eff}^{NP}(\mu^2)\eqno(2.8)$$
(the meaning of the quotes `` '' is clarified below) where it is assumed that each term in
eq.(2.7) satisfy separately a similar dispersion relation to eq.(2.2), e.g.:
$$\delta\alpha_s^{NP}(k^2) = k^2 \int_0^\infty{d\mu^2\over(\mu^2+k^2)^2}\
\delta\alpha_{eff}^{NP}(\mu^2)\eqno(2.9)$$ Furthermore, $\delta\alpha_s^{NP}$  generates
``non-perturbative''  power corrections :
\begin{eqnarray*}\ \ \ \ \ \ \ \ \ \ \ \ \ \ \ \ \ \ \ \delta
D_{NP}(Q^2)&=&\int_0^\infty{dk^2\over k^2}\
\delta\alpha_s^{NP}(k^2)\ \varphi\left(k^2\over Q^2\right)\ \ \ \ \ \ \ \ \ \ \ \ \ \ \ \ 
\ \ \ \ \ \ \ \ \ \ \ \ \ (2.10a)\\ &=&\int_0^\infty{d\mu^2\over\mu^2}\
\delta\alpha_{eff}^{NP}(\mu^2)\ \dot{{\cal F}}\left({\mu^2\over Q^2}\right)\ \ \ \ \ \ \ \ \
 \ \ \ \ \ \ \ \ \ \ \ \ \ \ \ \ \ \ \ (2.10b)\end{eqnarray*} The crucial further assumption of
[7] is that the ``non-perturbative'' contribution 
$\delta\alpha_s^{NP}(k^2)$, which reflects confinement physics, is essentially restricted to the
IR domain (in accordance with the OPE ideology of [20]), in order to comply with the requirement
that no power correction inconsistent with the ones expected from the OPE arises from the UV
behavior of $\delta\alpha_s^{NP}(k^2)$ . 

In fact, the precise meaning of
$``\alpha_s^{PT}(k^2)"$ was left open in [7]. It is one of the main purpose of this paper to fill
up this gap . Let us first note that both $``\alpha_s^{PT}(k^2)"$ and
$``\alpha_{eff}^{PT}(\mu^2)"$ should be IR regular: the former from the very assumption it
satisfies the dispersion relation, and thus cannot have any  Landau singularity, the latter
because any singularity at finite
$\mu^2$ (the dispersive variable) will make the dispersion relation and its output
$``\alpha_s^{PT}(k^2)"$ (hence $\delta\alpha_s^{NP}(k^2)$) ill-defined . It follows that none of
them can be given by such a simple form as (e.g.) the one-loop coupling. Nevertheless a simple
and attractive ansatz exists. I shall assume  that
$\alpha_s^{PT}(k^2)$ ({\em defined by a Borel sum }, see eq.(4.1) below) has no non-trivial IR
fixed point, but instead develops a Landau singularity on the space-like axis . Thus
$\alpha_s^{PT}(k^2)$ cannot satisfy the dispersion relation eq.(2.2), and the Landau singularity
has to be removed by hand~. This means one should understand $``\alpha_s^{PT}"$ in eq.(2.7) as
being:
$$\alpha_{s,reg}^{PT}=\alpha_s^{PT}+\delta\alpha_s^{PT}\eqno(2.11)$$   which differs from the
(Borel-summed)
$\alpha_s^{PT}$ by ``perturbative'' power corrections $\delta\alpha_s^{PT}$ which remove the
singularity . Upon insertion into eq.(2.1), $\alpha_{s,reg}^{PT}$ generates a ``regularized
perturbation theory ''[21,22] piece of $D(Q^2)$: 
$$D_{reg}^{PT}(Q^2)\equiv\int_{0}^\infty{dk^2\over k^2}\ \alpha_{s,reg}^{PT}(k^2)\
\varphi\left(k^2\over Q^2\right)\eqno(2.12)$$ which, as we sall see in section 6, contributes
new, ``perturbative'' type of power corrections . One can therefore write  $\alpha_s$ as:
$$\alpha_s=\alpha_s^{PT}+\delta\alpha_s\eqno(2.13)$$ where the total modification
$\delta\alpha_s$ comprises both a ``perturbative'' and a ``non-perturbative'' part:
$$\delta\alpha_s=\delta\alpha_s^{PT}+\delta\alpha_s^{NP}\eqno(2.14)$$ We shall see that, contrary
to $\delta\alpha_s^{NP}$, $\delta\alpha_s^{PT}$ is in general {\em not} restricted to low $k^2$,
and thus
$\delta\alpha_s\simeq\delta\alpha_s^{PT}\gg\delta\alpha_s^{NP}$ at large
$k^2$. Within these assumptions (which shall be relaxed in section 6.2), the determination of
$\alpha_{s,reg}^{PT}$ and $\delta\alpha_s^{PT}$ becomes a {\em physical question}, rather then a
matter of convention involved in the split eq.(2.14) between different components.  In the
time-like region ,  eq.(2.14) is paralleled by:
$$\alpha_{eff}(\mu^2)  =
\alpha_{eff}^{PT}(\mu^2)+\delta\alpha_{eff}^{PT}(\mu^2)+\delta\alpha_{eff}^{NP}(\mu^2)\eqno(2.15)$$
with:
$$\alpha_{s,reg}^{PT}(k^2) = k^2 \int_0^\infty{d\mu^2\over(\mu^2+k^2)^2}\
\left[\alpha_{eff}^{PT}(\mu^2)+\delta\alpha_{eff}^{PT}(\mu^2)\right]\eqno(2.16)$$ In the next
section, I investigate the simplest choice for $\delta\alpha_{eff}^{PT}$, namely
$\delta\alpha_{eff}^{PT}\equiv 0$, which implies:
$$D_{reg}^{PT}(Q^2)=\int_0^\infty{d\mu^2\over\mu^2}\ \alpha_{eff}^{PT}(\mu^2)\ \dot{{\cal
F}}({\mu^2\over Q^2})\eqno(2.17)$$

\section{Dispersive regularization}  A simple ansatz for $\alpha_{s,reg}^{PT}$  is illustrated by
the following example, which contains the essential ingredients of the general argument (and is
also relevant to ``large 
$\beta_0$'' QCD ). Consider the ``minimal regularization'' of the one loop coupling , obtained by
just removing the Landau pole at
$k^2=\Lambda^2$, i.e. define:

\begin{eqnarray*}\ \ \ \ \ \ \ \ \ \ \ \ \ \ \ \ \ \ \ \ \ \ \ \alpha_{s,reg}^{PT}(k^2)&
\equiv&{1\over\beta_0\ln(k^2/\Lambda^2)}- {1\over\beta_0}{1\over{k^2\over
\Lambda^2}-1 }\\ &\equiv&\alpha_s^{PT}(k^2) + \delta\alpha_s^{PT}(k^2)\ \ \ \ \ \ \ \ \ \ \ \ \ \
\ \ 
\ \ \ \ \ \ \ \ \ \ \ \ \ \ \ \ \ \ \ (3.1)\end{eqnarray*} ($\beta_0>0$). The resulting
$\alpha_{s,reg}^{PT}(k^2)$ is analytic in the complex $k^2$ plane , with a cut on the negative
$k^2$ axis (the time-like region in our notation), and satisfies the dispersion relation
eq.(2.2). The simple, but crucial observation which makes the above model interesting is that the
corresponding time-like discontinuity is {\em entirely} given by that  of the one loop coupling:
$$\rho_{reg}^{PT}(\mu^2) = \rho_{PT}(\mu^2) \equiv -\ {1\over\beta_0}\ \frac {1} {\ln^2 {\mu^2
\over
\Lambda^2}+\pi^2}\eqno(3.2)$$ One notices that $\rho_{PT}(\mu^2)$ is {\em continuous, finite and
negative} in the whole range \\ $0<\mu^2<\infty$  of the dispersive variable, and vanishes both
for $\mu^2 \rightarrow
\infty$ {\em and} $\mu^2 \rightarrow 0$~. It follows the corresponding
$\alpha_{eff,reg}^{PT}$ coincides with
$\alpha_{eff}^{PT}$ (i.e. $\delta\alpha_{eff}^{PT}\equiv 0$) and is obtained by substituting
$\rho$ with $\rho_{PT}$ in eq.(2.4), which gives:
\begin{eqnarray*}\ \ \ \ \ \ \ \ \ \
\ \ \ \ \ \ \alpha_{eff}^{PT}(\mu^2)&=&\frac{1}{\pi\beta_0}\left[{\pi\over
2}-\arctan\left(\frac{1}{\pi}\ln{\mu^2
\over\Lambda^2}\right)\right]\\ &\equiv&\frac{1}{\pi\beta_0}\arctan\left(\frac{\pi}{\ln{\mu^2
\over\Lambda^2}}\right)\ \ \ \ \ \ \ \ \ \ (\mu^2>\Lambda^2)\ \ \ \ \ \ \ \
\ \ \ \ \ \ \ \ \ \ \ \ (3.3a)\\ &\equiv&\frac{1}{\pi\beta_0}\arctan\left(\frac{\pi}{\ln{\mu^2
\over\Lambda^2}}\right)+\frac{1}{\beta_0}\ \ \ \ (0<\mu^2<\Lambda^2)\ \ \ \ \ \ \ \
\ \ \ \ \ \ (3.3b)\end{eqnarray*}  (this coupling was introduced previously in [6] (see also
[13])  with a somewhat different motivation)~. The ``effective coupling''
$\alpha_{eff}^{PT}(\mu^2)$ is IR finite and satisfies the RG equation:
$${d\alpha_{eff}^{PT}\over d\ln \mu^2}=-{1 \over
\pi^2\beta_0}\sin^2(\pi\beta_0\alpha_{eff}^{PT})\eqno(3.4)$$ It therefore increases from $0$ to
the IR fixed point value $1/\beta_0$ as $\mu^2$ is decreased from
$\infty$ to $0$. The corresponding dispersively generated  $\alpha_{s,reg}^{PT}(k^2)$ differs
from the one-loop coupling $\alpha_s^{PT}(k^2)$ by power corrections (see eq.(3.1)), and
approaches also $1/\beta_0$ as $k^2\rightarrow 0$. There is thus non-commutativity of resummation
(of e.g. the series obtained when
$\alpha_{eff}^{PT}(\mu^2)$ is expanded in powers  $\alpha_s^{PT}(k^2)$) and integration under the
dispersive integral eq.(2.2b), reflecting the non-trivial IR fixed point of
$\alpha_{eff}^{PT}(\mu^2)$: this is an example of the general phenomenon discussed in [23,24].

The features observed for the one loop coupling, namely, negative definite 
$\rho_{PT}(\mu^2)$~, vanishing of
$\rho_{PT}(\mu^2=0)$ , and IR finitness  of $\alpha_{eff}^{PT}(\mu^2)$  are likely to remain true
at any number of loops. Indeed, it seems  reasonnable to  assume that the only singularity of
$\alpha_s^{PT}(k^2)$ on the first sheet of the cut complex $k^2$ plane is the space-like Landau
singularity, 
 and in particular that $\alpha_s^{PT}(k^2)$, hence its discontinuity $\rho_{PT}(\mu^2)$, remain
finite on the time-like axis. If there is no time-like singularity ( and no non-perturbative
thresholds are expected in the {\em perturbative} part of the coupling),  the discontinuity
should be {\em continuous} for $0<\mu^2<\infty$   and cannot  change sign  without going through
{\em real} values of
$\alpha_s^{PT}$. But real values are in general not compatible with the constant $i\pi$ imaginary
part acquired by $\ln k^2$ upon analytic continuation from the space-like to the time-like
region. For instance, assume $\alpha_s^{PT}$ satisfies the 2-loop RG equation:
$${d\alpha_s^{PT}\over d\ln k^2} = -\beta_0 (\alpha_s^{PT})^2 - \beta_1
(\alpha_s^{PT})^3\eqno(3.5)$$ whose solution is:
$$\beta_0 \ln {k^2\over\Lambda^2} = {1\over a_s}+\lambda \ln a_s \eqno(3.6)$$   with
$a_s=\alpha_s/ (1+\lambda \alpha_s)$ and $\lambda =\beta_1/\beta_0$. One finds, upon going to the
time-like region~, setting $\ln k^2 = \ln \mu^2 + i\pi$ $(\mu^2>0)$, $a_s=\mid a_s\mid
\exp i\theta$ and taking the imaginary part:
$$\pi \beta_0 = {\sin \theta \over \mid a_s\mid}- \lambda \theta\eqno(3.7)$$   where I assumed
$\theta >0$ (  the sign is appropriate for RG trajectories in the domain of attraction of the {\em
trivial} UV fixed point). It is clear that $\sin \theta$ can vanish (hence $Im\ \alpha_s$ can
change sign) at {\em finite} $\mu^2$ only in the special cases  where
$n\lambda =-\beta_0$ ($n$ positive integer), which are excluded anyway since $\lambda >0$. 
Assuming the discontinuity indeed does not change sign, asymptotic freedom fix it to be
negative~.  Furthermore, {\em if one assumes} that $\alpha_s^{PT}(k^2)$ approaches the {\em
trivial} IR fixed point for
$k^2\rightarrow 0$, i.e. that
$\alpha_s^{PT}(k^2=0)=0$, then
$\alpha_{eff}^{PT}(\mu^2)$ is IR finite. Indeed one gets (see eq.(2.4)):
\begin{eqnarray*}\ \ \ \
\ \ \ \ \ \ \ \ \ \ \ \ \ \
\alpha_{eff}^{PT}(\mu^2)&=&-\int_{\mu^2}^\infty{d\mu'^2\over\mu'^2}\
\rho_{PT}(\mu'^2)\ \ \ \ \ \ \ \
\ \ \ \ \ \ \ \ \ \
\ \ \ \ \ \ \ \ \ \ \ \ \ \ \ \ \ \ \ \ \ \ \ (3.8a)\\ &=&\int_0^{\mu^2}{d\mu'^2\over\mu'^2}\
\rho_{PT}(\mu'^2)+\alpha_{eff\mid IR}^{PT}\ \ \ \ \ \ \ \
\ \ \ \ \ \ \ \ \ \
\ \ \ \ \ \ \ \ \ \ \ \ \ (3.8b)\end{eqnarray*} where:
$$\alpha_{eff\mid IR}^{PT}\equiv\alpha_{eff}^{PT}(\mu^2=0)=-\int_0^\infty{d\mu^2\over\mu^2}\
\rho_{PT}(\mu^2)\eqno(3.9)$$ and the integrals converge at $\mu^2=0$ since
$\rho_{PT}$  vanishes there.

\noindent It is therefore natural to {\em define}
$\alpha_{s,reg}^{PT}$ through the dispersion relation:
\begin{eqnarray*}\ \ \ \ \
\ \ \ \ \ \ \ \ \
\ \ \ \ \ \ \ \ \  \ \ \ \ \ \ \ \ \alpha_{s,reg}^{PT}(k^2)
&=&-\int_0^\infty{d\mu^2\over\mu^2+k^2}\
\rho_{PT}(\mu^2)\ \ \ \ \ \ \ \
\ \ \ \ \ \ \ \ \ \ \ \ \ \ \ \ \ (3.10a)\\ &=& k^2\int_0^\infty{d\mu^2\over(\mu^2+k^2)^2}\
\alpha_{eff}^{PT}(\mu^2)\ \ \ \ \ \ \ \
\ \ \ \ \ \ \ \ \ \ \ \ (3.10b)\end{eqnarray*}    i.e. take $\delta\alpha_{eff}^{PT}\equiv 0$ in
eq.(2.15). Although this suggestion is new in the present context, I realized while writing this
article that the resulting ``dispersive regularization'' of the Landau singularity has actually
been proposed [10,11] almost 40 years ago in QED, and has been revived recently in QCD [12,13].
Since $\alpha_{eff}^{PT}(\mu^2)$ has a non-trivial   IR fixed-point,  
$\alpha_{s,reg}^{PT}(k^2)$  differs [23,24] from 
$\alpha_s^{PT}(k^2)$ by an infinite set of ``dispersively generated'' power corrections
$\delta\alpha_s^{PT}(k^2)$, of perturbative origin, which remove the Landau singularity. It also
follows from eq.(3.9)-(3.10) that $\alpha_{s,reg}^{PT}(k^2=0) =
\alpha_{eff\mid IR}^{PT}$~.

For a more general example then the one-loop coupling, assume the Landau singularity is a cut
starting in the space-like region at
$k^2=\Lambda^2$, and there is no further singularity in the first sheet of the complex $k^2$
plane. One can then write the dispersion relation:
$$\alpha_s^{PT}(k^2) = -\int_{-\Lambda^2}^\infty{d\mu^2\over\mu^2+k^2}\
\rho_{PT}(\mu^2)\eqno(3.11)$$ Splitting-off the space-like discontinuity from $-\Lambda^2$ to $0$
one immediately gets eq.(2.11) with $\alpha_{s,reg}^{PT}$ as in eq.(3.10a) and:
$$\delta\alpha_s^{PT}(k^2)=
\int_{-\Lambda^2}^0{d\mu^2\over\mu^2+k^2}\ \rho_{PT}(\mu^2)\eqno(3.12)$$ whose large $k^2$
expansion is:
$$\delta\alpha_s^{PT}(k^2)=-\sum_{n=1}^\infty b_n^{PT} \left(-\ {\Lambda^2\over
k^2}\right)^n\eqno(3.13)$$  where the $b_n^{PT}$'s are real numbers : 
$$b_n^{PT} = \int_{-\Lambda^2}^0{d\mu^2\over\mu^2}\left({\mu^2\over\Lambda^2}\right)^n
\rho_{PT}(\mu^2)\eqno(3.14)$$ Furthermore ,  {\em if one assumes} that $\alpha_s^{PT}(k^2=0)$
vanishes,   we have (setting $k^2=0$ in eq.(3.12)):
\begin{eqnarray*}\ \ \ \ \ \ \ \
\alpha_{s,reg}^{PT}(k^2=0)=\delta\alpha_s^{PT}(k^2=0)&=&\int_{-\Lambda^2}^0 {d\mu^2\over\mu^2}\
\rho_{PT}(\mu^2)\equiv b_0^{PT}\\ &=&-\int_0^\infty{d\mu^2\over \mu^2}\
\rho_{PT}(\mu^2)\equiv\alpha_{eff\mid IR}^{PT}\ \ \ \ \ \ \ \ \ \ (3.15)\end{eqnarray*}

\section{Borel transform techniques}  A more specific method to obtain the ``perturbative power
corrections'' $\delta\alpha_s^{PT}(k^2)$  makes use of the ``RS invariant Borel transform''
[25-27] , i.e. I shall assume that $\alpha_s^{PT}$ is given by: 
$$\alpha_s^{PT}(k^2)=\int_0^\infty dz \exp\left(-z\beta_0\ln{k^2 \over
\Lambda^2}\right)\ \tilde{\alpha}_s(z) \ \ \ \ \ (k^2>\Lambda^2)\eqno(4.1)$$ where it is
convenient to choose $\Lambda$ as the Landau singularity of $\alpha_s^{PT}$  . The ``RS invariant
Borel transform''
$\tilde{\alpha}_s(z)$ is simply related to the ordinary transform   (and coincides with the
latter in the ``'t Hooft scheme'' (eq.(3.5)) if
$\beta_1=0$); for the one-loop coupling, $\tilde{\alpha}_s(z)\equiv 1$. In this section, I assume
$\tilde{\alpha}_s(z)$ has no IR renormalons singularities, so that eq.(4.1)  defines
unambiguously $\alpha_s^{PT}(k^2)$ for $k^2>\Lambda^2$. Taking the time-like discontinuity of
eq.(4.1), one gets:
$$\rho_{PT}(\mu^2)=\int_0^\infty dz \exp\left(-z\beta_0\ln{\mu^2 \over
\Lambda^2}\right)\ \tilde{\rho}(z) \ \ \ \ \ (\mu^2>\Lambda^2)\eqno(4.2)$$  with: 
$$\tilde{\rho}(z)=-\ {1\over\pi}\sin(\pi\beta_0 z)\ \tilde{\alpha}_s(z)\eqno(4.3)$$ Hence, using
eq.(3.8a):
$$\alpha_{eff}^{PT}(\mu^2)=\int_0^\infty dz
\exp\left(-z\beta_0\ln{\mu^2 \over
\Lambda^2}\right)\ \tilde{\alpha}_{eff}(z) \ \ \ \ \ \ \ \ \ \ \ \ \ \ \ \ \
(\mu^2>\Lambda^2)\eqno(4.4)$$  where $\tilde{\alpha}_{eff}(z)$ is the RS invariant Borel
transform  of $\alpha_{eff}^{PT}(\mu^2)$:
$$\tilde{\alpha}_{eff}(z)=-\ {\tilde{\rho}(z)\over z\beta_0 }={\sin(\pi\beta_0 z)\over\pi\beta_0 z
}\ \tilde{\alpha}_s(z)\eqno(4.5)$$ Eq.(4.5) is  just the relation between the ``modified Borel
transforms'' of absorptive and dispersive parts first obtained in [27].  The oscillations of the
$\sin(\pi\beta_0 z)$ factor in eq.(4.5) account for the  absence of Landau singularity of
$\alpha_{eff}^{PT}(\mu^2)$ (despite its  presence in
$\alpha_s^{PT}(k^2)$). For instance in the case of the one-loop coupling where
$\tilde{\alpha}_s(z)\equiv 1$ ,  eq.(4.4) and (4.5))  reproduce eq.(3.3a). Note the alternative
ansatz $\tilde{\alpha}_{eff}(z)\equiv 1$, i.e. assuming
$\alpha_{eff}^{PT}(\mu^2)$ itself is the one-loop coupling (hence singular at
$\mu^2=\Lambda^2$), cannot give a consistent answer upon insertion into the dispersion  relation~
(and would imply renormalons in $\tilde{\alpha}_s(z)$!). Furthermore, the assumption    that
$\alpha_{eff}^{PT}(\mu^2)$  is IR regular  explains [23,24] that the right hand side of
eq.(3.10b) may differ from its Borel sum $\alpha_s^{PT}$ by power corrections, and also that
$\tilde{\alpha}_s(z)$ has no IR renormalons generated by the low energy part of the dispersive
integral in eq.(3.10b) (which would reflect the ambiguity of integrating over an IR singular
$\alpha_{eff}^{PT}$). 

\noindent An alternative way to derive eq.(4.5) starts from eq.(3.10b), where one freely replaces
$\alpha_{eff}^{PT}(\mu^2)$ by its Borel representation eq.(4.4)  inside the dispersive integral
(although this is not justified for
$\mu^2<\Lambda^2$!). Interchanging the order of the
$z$ and
$\mu^2$ integrations yields $\alpha_s^{PT}(k^2)$ ( not
$\alpha_{s,reg}^{PT}(k^2)$!) as in eq.(4.1) , with:
\begin{eqnarray*}\ \ \ \ \ \ \ \ \ \tilde{\alpha}_s(z) & = & \tilde{\alpha}_{eff}(z)\  k^2\int_0^
\infty {d\mu^2\over(\mu^2+k^2)^2}\ \exp\left(-z\beta_0\ln{\mu^2\over k^2 }\right)\\ &=&
\tilde{\alpha}_{eff}(z)\  {\pi\beta_0 z\over\sin(\pi\beta_0 z)}\ \ \ \ \ \ \ \ \ \
\ \ \ \ \ \ \ \ \ \
\ \ \ \ \ \ \ \ \ \
\ \ \ \ \ \ \ \ \ \
\ \ \ \ \ \ \ \ \ \ \ \ \ \ \ \ \ (4.6)\end{eqnarray*} 

To derive the power corrections in $\delta\alpha_s^{PT}$ it is convenient (although not absolutely
necessary, see Appendix A2) to first split the dispersive integral eq.(3.10b) at $\mu^2=k^2$:
$$\alpha_{s,reg}^{PT}(k^2)  = k^2\int_0^{k^2}{d\mu^2\over(\mu^2+k^2)^2}\
\alpha_{eff}^{PT}(\mu^2)+k^2\int_{k^2}^\infty{d\mu^2\over (\mu^2+k^2)^2}\
\alpha_{eff}^{PT}(\mu^2)\eqno(4.7)$$ The second integral contributes only to the Borel sum
eq.(4.1), and not to the power corrections , since one can use the Borel representation of
$\alpha_{eff}^{PT}$ (eq.(4.4)) there (taking
$k^2>\Lambda^2$). On the other hand, expanding the dispersive kernel in the first integral in
inverse powers of
$k^2$ one gets:
$$k^2\int_0^{k^2}{d\mu^2\over(\mu^2+k^2)^2}\
\alpha_{eff}^{PT}(\mu^2)=\sum_{n=1}^\infty (-1)^{n+1}I_n(k^2)\eqno(4.8)$$ with
$$I_n(k^2)\equiv \int_0^{k^2}n{d\mu^2\over\mu^2}\left({\mu^2\over k^2}\right)^n
\alpha_{eff}^{PT}(\mu^2)\eqno(4.9)$$

The $I_n(k^2)$'s are standard IR renormalons integrals. If the coupling
$\alpha_{eff}^{PT}(\mu^2)$ has a non-trivial IR fixed point,  they differ [23,24]  from their
corresponding Borel sums
$I_n^{PT}(k^2)$ by a power correction. Putting:
$$I_n^{PT}(k^2) = \int_0^\infty dz \exp\left(-z\beta_0\ln{k^2 \over
\Lambda^2}\right)\ \tilde{I}_n(z) \ \ \ \ \ (k^2>\Lambda^2)\eqno(4.10)$$ one gets:
\begin{eqnarray*}\ \ \ \ \ \ \ \ \ \ \ \ \ \ \ \ \ \ \ \ \ \ \ \tilde{I}_n(z) & = &
\tilde{\alpha}_{eff} (z)\ 
\int_0^{k^2}n{d\mu^2\over
\mu^2}\
\left({\mu^2\over k^2}\right)^n\
\exp\left(-z\beta_0\ln{\mu^2\over k^2 }\right)\\ &=& \tilde{\alpha}_{eff}(z)\
\frac{1}{1-\frac{z}{z_n}}\ \ \ \ \ \ \ \ \ \
\ \ \ \ \ \ \ \ \ \
(z_n={n\over \beta_0})
\ \ \ \ \ \ \ \ \ \
\ \ \ \ (4.11)\end{eqnarray*} whereas:
$$I_n(k^2)=I_n^{PT}(k^2)+b_n^{PT} \left({\Lambda^2\over k^2}\right)^n\eqno(4.12) $$ with [23]:
$$b_n^{PT}=I_n - I_n^{PT}\eqno(4.13)$$ where:
$$I_n\equiv I_n(k^2=\Lambda^2)=\int_0^{\Lambda^2}n{d\mu^2\over\mu^2}\left({\mu^2\over
\Lambda^2}\right)^n
\alpha_{eff}^{PT}(\mu^2)\eqno(4.14)$$ and:
$$I_n^{PT}\equiv I_n^{PT}(k^2=\Lambda^2)=\int_0^\infty dz\  
\tilde{\alpha}_{eff}(z)\
\frac{1}{1-\frac{z}{z_n}}\eqno(4.15)$$

\noindent To derive these results (see also section 7.1) one splits the integral in eq.(4.9) at
$\mu^2=\Lambda^2$. The low energy integral is just $I_n\left({\Lambda^2\over k^2}\right)^n$,
whereas in the high energy integral, one can use the Borel representation eq.(4.4) of
$\alpha_{eff}^{PT}$ to get  
$I_n^{PT}(k^2)-I_n^{PT} \left({\Lambda^2\over k^2}\right)^n $ . Adding the two pieces yields
eq.(4.12). Note that, provided
$\tilde{\alpha}_s(z)$ has no renormalons, so does $\tilde{I}_n(z)$, since  the zeroes of
$\tilde{\alpha}_{eff}(z)$ (eq.(4.5)) sit precisely at the would-be renormalons positions when $n$
is an integer. Consequently, the power correction in eq.(4.12) and the constants $b_n^{PT}$ are
{\em real} and {\em unambiguous}, but nevertheless $I_n(k^2)$ differs from its {\em well defined}
Borel sum
$I_n^{PT}(k^2)$ : this is an example of the phenomenon pointed out in [24].  Since the power
corrections in $\alpha_{s,reg}^{PT}(k^2)$ are given by those in the
$I_n(k^2)$'s, one recovers eq.(3.13) from eq.(4.8).

It is instructive to rederive the result eq.(3.1) for the regularized one-loop coupling with the
above method. In this case, not only $I_n^{PT}$, but also the constants $I_n$ and
$b_n^{PT}$ can be computed from the Borel transform $\tilde{\alpha}_{eff}(z)={\sin(\pi\beta_0
z)\over\pi\beta_0 z}$~, since
$\alpha_s^{PT}(k^2)$    satisfies for $0<k^2<\Lambda^2$ the
${\em z<0}$ Borel representation : 
$$\alpha_s^{PT}(k^2) =-\int_{-\infty}^0 dz \exp\left(-z\beta_0\ln{k^2 \over
\Lambda^2}\right)\ \tilde{\alpha}_s(z) \ \ \ (0<k^2<\Lambda^2)\eqno(4.16) $$ with $\
\tilde{\alpha}_s(z)\equiv 1$. Taking the time-like discontinuity of eq.(4.16), one finds~:
$$\rho_{PT}(\mu^2)=-\int_{-\infty}^0 dz \exp\left(-z\beta_0\ln{\mu^2
\over\Lambda^2}\right)\ \tilde{\rho}(z) \
\  \ (0<\mu^2<\Lambda^2)\eqno(4.17) $$ with $\tilde{\rho}(z)=-{1\over\pi}\sin(\pi\beta_0 z)$. 
From eq.(3.8b) one  deduces:
\begin{eqnarray*}\ \ \ \ \ \ \ \ \
\alpha_{eff}^{PT}(\mu^2)&=&\alpha_{eff\mid IR}^{PT} - \int_{-\infty}^0 dz
\exp\left(-z\beta_0\ln{\mu^2
\over
\Lambda^2}\right)\ \tilde{\alpha}_{eff}(z)\ \ \ \ \ \ \ \ \ \ \ \ \ \ \ \ \ \ \ \ (4.18) \\
 & &\ \ \ \ \ \ \ \ \ \ \ \ \ \ \ \ \ \ \ \ \ \ \ \ \ \ \
\ \ \ \ \ \ \ \ \ \ \ \ \ \ \ \ \ \ \ \ \ \ \ \ \ \ \ \
\ \ \ \ (0<\mu^2<\Lambda^2)\end{eqnarray*} Note that eq.(4.4) and (4.18) imply (for
$\mu^2=\Lambda^2$):
$$\alpha_{eff\mid IR}^{PT}=\int_{-\infty}^\infty  dz\ \tilde{\alpha}_{eff}(z)={1\over
\beta_0}\eqno(4.19)$$ Inserting eq.(4.18) into eq.(4.14) , one gets:
$$I_n=\alpha_{eff\mid IR}^{PT}-\bar{I}_n^{PT} \eqno(4.20) $$ with
$$\bar{I}_n^{PT}=\int_{-\infty}^0 dz\ \tilde{\alpha}_{eff}(z)\
\frac{1}{1-\frac{z}{z_n}}
\eqno(4.21) $$ hence (eq.(4.13)):
$$b_n^{PT}=\alpha_{eff\mid IR}^{PT} - \left(I_n^{PT}+\bar{I}_n^{PT}\right)\eqno(4.22) $$ which,
upon substitution into eq.(3.13) yields:
$$\delta\alpha_s^{PT}(k^2)=\alpha_{eff\mid IR}^{PT}\ \frac {{\Lambda^2\over k^2}}
{1+{\Lambda^2\over k^2}}+\sum_{n=1}^\infty \left(I_n^{PT}+\bar{I}_n^{PT}\right) 
\left(-\ {\Lambda^2\over k^2}\right)^n\eqno(4.23) $$ But:
\begin{eqnarray*}\ \ \ \ \ \ \ \ \ \ \ I_n^{PT}+\bar{I}_n^{PT}&=&\int_{-\infty}^{\infty} dz\
{\sin(\pi\beta_0 z)\over\pi\beta_0 z }\ \frac{1}{1-\frac{z}{z_n}}\\  &=&{1\over\beta_0} (1-(-1)^n
)=\left\{\begin{array}{lc}0&(n\ even)\\{2\over\beta_0}&(n\ odd)\end{array}\right.\ \ \ \ \ \ \ \
\ \ \ \ \ \ \ \ \ \ \ \ \ \ \ \ \ (4.24)\end{eqnarray*} which gives, since $\alpha_{eff\mid
IR}^{PT}={1\over \beta_0}$  (the latter value is  actually universal [12] and holds beyond one
loop within some assumptions, see Appendix A):
$$b_n^{PT}=(-1)^n {1\over \beta_0}\eqno(4.25)$$ A similar  method could  be applied to the
two-loop coupling, where [25]: 
$$\tilde{\alpha}_s(z)=\exp(\lambda z \ln\lambda z ) {\exp(-\lambda z )\over
\Gamma(1+\lambda z)}\eqno(4.26)$$ Here one should take into account the fact that
$\tilde{\alpha}_s(z)$ is {\em complex} for
$z<0$, since the Landau singularity is a cut rather then a pole in this case.

\underline {Infrared fixed point case}: No ``regularization'' is needed if one assumes that the
Borel-summed $\alpha_s^{PT}(k^2)$ satisfies by itself the dispersion relation eq.(2.2).   A
simple example of such a coupling, relevant to the ``small
$\beta_0$''limit of QCD [28], is provided by the two loop beta function (eq.(3.5)) with
${\beta_1\over\beta_0}<0$, which has an IR fixed point $\alpha_{s\mid IR}^{PT}=-{\beta_0\over
\beta_1}$. As noted by Uraltsev [29], for large
$\beta_0$, more precisely if $\beta_1<0$ but
${\beta_1\over\beta_0}+\beta_0>0$, there are complex Landau singularities, which move to the
second sheet when $\beta_0$ is decreased and ${\beta_1\over\beta_0}+\beta_0<0$. In the latter
case ( barring further finite singularities ) $\alpha_{s,reg}^{PT}\equiv \alpha_s^{PT}$ and  all
power corrections  vanish ( i.e. $I_n=I_n^{PT}$)\footnote{If however 
$\tilde{\alpha}_s(z)$ has renormalons (see next section),  $\alpha_s^{PT}$ (defined by a Borel
sum) can of course never  satisfy  eq.(2.2).}\ .

\section{Renormalons in $\tilde{\alpha}_s(z)$} Up to now, I  assumed that $\tilde{\alpha}_s(z)$
has no renormalons. In reality  , this is likely not to be the case, since $\alpha_s$ is a
physical coupling analoguous to the Gell-Mann Low effective charge in QED, which probably does
have renormalons. The general idea of constructing the regularized perturbative coupling through
a dispersion relation from the discontinuity of the  perturbative coupling still applies, but
there are  conceptual as well as technical complications, since in this case even the latter 
cannot be defined by a Borel sum as in eq.(4.2) without additionnal prescription. I shall limit
myself to the following simple example: assume the ``physical'' QCD coupling
$\bar{\alpha}_s^{PT}(k^2)$ (I use a superscript ``bar'' to avoid confusion with the ``t'Hooft
coupling'' below) is given by the standard IR renormalon integral:
$$\bar{\alpha}_s^{PT}(k^2)=\int_0^{k^2} n \frac{dk'^2}{k'^2}
\left(\frac{k'^2}{k^2}
\right)^n
\alpha_s^{PT}(k'^2)\eqno(5.1)$$ where $\alpha_s^{PT}(k^2)$ is the ``t'Hooft coupling'' of
eq.(3.5). It has been shown in [23] that  $\bar{\alpha}_s^{PT}(k^2)$ satisfies the ({\em
ordinary}) Borel representation with respect to $\alpha\equiv \alpha_s^{PT}(k^2)$~:
$$\bar{\alpha}_s^{PT}(k^2)=\int_0^\infty dz\ exp\left(- \frac{z}{\alpha}\right) 
\frac{exp\left(-{\beta_1\over \beta_0} z\right)}{\left(1 - \frac{z}{z_n}
\right)^{1+\delta}}\eqno(5.2)$$ with $\delta={\beta_1\over \beta_0} z_n$. Note that the {\em
standard} Borel transform singularity at $z=z_n$ is a cut [30] if $\beta_1\neq 0$. It follows
from [27] that the corresponding {\em modified, ``RS invariant''} Borel transform
$\tilde{\bar{\alpha}}_s(z)$  has a simple pole singularity at $z=z_n$. For instance, if
$\beta_1 = 0$ (in which case the ordinary and modified Borel transforms coincide), one has
$\tilde{\bar{\alpha}}_s(z) = \frac{1}{1-\frac{z}{z_n}}$, hence (eq.(4.5)) 
\ $\tilde{\bar{\alpha}}_{eff}(z) = {\sin(\pi\beta_0 z)\over\pi\beta_0 z}\
\frac{1}{1-\frac{z}{z_n}}$, which coincides with the one-loop
$\tilde{I}_n(z)$ ( see eq.(4.11)). The latter fact is not accidental, as one can  show [18] (see
section 7) that the time-like discontinuity of
$\bar{\alpha}_s^{PT}(k^2)$ in eq.(5.1) (properly extended (section 7) to the complex $k^2$ plane)
is given by the analoguous integral \\( which implies a peculiar definition of the Borel sum in
eq.(5.2) to be consistent with eq.(5.1)) :
$$\bar{\rho}_{reg}^{PT}(\mu^2) = \int_0^{\mu^2} n \frac{d\mu'^2}{\mu'^2}
\left(\frac{\mu'^2}{\mu^2}\right)^n\ \rho_{PT}(\mu'^2)\eqno(5.3)$$ hence: 
$$\bar{\alpha}_{eff,reg}^{PT}(\mu^2) = \int_0^{\mu^2} n \frac{d\mu'^2}{\mu'^2}
\left(\frac{\mu'^2}{\mu^2}\right)^n\ \alpha_{eff}^{PT}(\mu'^2)\eqno(5.4)$$   (where $\rho_{PT}$
and $\alpha_{eff}^{PT}$ are the corresponding quantities for
$\alpha_s^{PT}$). Eq.(5.4) shows that
$\bar{\alpha}_{eff,reg}^{PT}(\mu^2)$ coincides with
$I_n(k^2=\mu^2)$ (eq.(4.9)), and differs from the corresponding Borel sum
$\bar{\alpha}_{eff}^{PT}(\mu^2) = \int_0^\infty dz
\exp\left(-z\beta_0\ln{\mu^2 \over
\Lambda^2}\right)\ \tilde{\bar{\alpha}}_{eff}(z)$ by a power correction, as we have seen in 
section 4. One can also show (section 7) that the $\bar{\alpha}_{s,reg}^{PT}(k^2)$ following from
the dispersive regularization procedure:
\begin{eqnarray*}\ \ \ \ \
\ \ \ \ \ \ \ \ \
\ \ \ \ \ \ \ \ \  \ \ \ \ \ \ \ \ \bar{\alpha}_{s,reg}^{PT}(k^2)
&=&-\int_0^\infty{d\mu^2\over\mu^2+k^2}\
\bar{\rho}_{reg}^{PT}(\mu^2)\\ &=& k^2\int_0^\infty{d\mu^2\over(\mu^2+k^2)^2}\
\bar{\alpha}_{eff,reg}^{PT}(\mu^2)\ \ \ \ \ \ \ \
\ \ \ \ \ \ \ \ (5.5)\end{eqnarray*} is simply given by the similar formula: 
$$\bar{\alpha}_{s,reg}^{PT}(k^2) = \int_0^{k^2} n \frac{dk'^2}{k'^2}
\left(\frac{k'^2}{k^2}
\right)^n
\alpha_{s,reg}^{PT}(k'^2)\eqno(5.6)$$ (which displays a certain ``self-consistency'' of the
procedure).

\section{``Perturbative'' and ``non-perturbative''power corrections}  Upon insertion of
eq.(2.13), the representation eq.(2.1) for the Euclidean observable $D(Q^2)$  , can be split,  
as we have seen, into a  ``regularized perturbation theory'' [21,22] and a `` genuine
non-perturbative'' [7] piece :
\begin{eqnarray*}\ \ \ \ \ \ \ \ \ \ \ \ \ \ \ \ D(Q^2)& =&\int_{0}^\infty{dk^2\over k^2}\
\alpha_{s,reg}^{PT}(k^2)\ \varphi\left(k^2\over Q^2\right) + \int_{0}^\infty{dk^2\over k^2}\
\delta\alpha_s^{NP}(k^2)\ \varphi\left(k^2\over Q^2\right)\\ &\equiv&D_{reg}^{PT}(Q^2)+\delta
D_{NP}(Q^2)\ \ \ \ \ \ \ \ \ \ \ \ \ \ \ \ \ \ \ \ \ \ \ \
\ \ \ \ \ \ \ \ \ \ \ \ \ \ \ \ \ \ \ (6.1)\end{eqnarray*} Each of these contributions,  which I
consider  in turn, generates power corrections .

\subsection{``Perturbative'' power corrections} The ``regularized  perturbation theory'' piece
$D_{reg}^{PT}$ may be decomposed, following eq.(2.11), as:
$$ D_{reg}^{PT}(Q^2) = D_{PT}(Q^2)+\delta D_{PT}(Q^2)\eqno(6.2)$$ with:
$$D_{PT}(Q^2)=\int_{0}^\infty{dk^2\over k^2}\ \alpha_s^{PT}(k^2)\ \varphi\left(k^2\over
Q^2\right)\eqno(6.3)$$ and:
$$\delta D_{PT}(Q^2)=\int_{0}^\infty{dk^2\over k^2}\ \delta\alpha_s^{PT}(k^2)\
\varphi\left(k^2\over Q^2\right)\eqno(6.4)$$ {\em Provided} $\alpha_s^{PT}(k^2)$ has no
non-trivial IR fixed point, and satisfies for small enough
$k^2$   the
$z<0$ Borel representation eq.(4.16),
$D_{PT}(Q^2)$ can be identified [23]  to the  perturbation theory Borel sum:
$$D_{PT}(Q^2)=\int_0^\infty dz \exp\left(-z\beta_0\ln{Q^2 \over
\Lambda^2}\right)\ \tilde{D}(z)\ \ \ \ \ \ \ \ \ \ \ \ \ (Q^2>\Lambda^2)\eqno(6.5a)$$   with:
$$\tilde{D}(z)= \tilde{\alpha}_s(z)\  \int_0^\infty{dk^2\over k^2}\ \varphi\left(k^2\over
Q^2\right)\ \exp\left(-z\beta_0\ln{k^2\over Q^2 }\right)\eqno(6.5b)$$ Eq.(6.5) is obtained in
practice by freely using the $k^2>\Lambda^2$ Borel representation of
$\alpha_s^{PT}(k^2)$ (eq.(4.1)) in  eq.(6.3), and  permutting the $k^2$ and $z$ integrations.
(This procedure can be justified [23] by splitting the integral in eq.(6.3) at $k^2=Q^2$. The
$k^2>Q^2$ piece poses no problem. The $k^2<Q^2$ piece is defined by analytic continuation from 
the low $Q^2$ region, where one can use the $z<0$ Borel representation eq.(4.16) of the coupling
to derive a $z<0$ Borel representation ).\\
$D_{reg}^{PT}(Q^2)$ differs from the Borel sum
$D_{PT}(Q^2)$ by ``perturbative'' power corrections $\delta D_{PT}(Q^2)$. These corrections are
expected, the presence of the Landau singularity of $\alpha_s^{PT}(k^2)$ in the integration range
making $D_{PT}$ ill-defined (an ambiguity  reflected in the usual way through the presence of IR
renormalons at $z=z_n>0$ in $\tilde{D}(z)$), whereas
$D_{reg}^{PT}$ is unambiguous. However, the important  point 
 is that
$\delta\alpha_s^{PT}(k^2)$ is {\em only moderately suppressed} at high $k^2$, i.e.   most likely
(eq.(3.13))
$\delta\alpha_s^{PT}(k^2) = {\cal O}(\Lambda^2/k^2)$ for $k^2\gg\Lambda^2$. Consequently,  
 $\delta D_{PT}(Q^2)$ will also get (apart from the ambiguous IR contributions which must be
present to cancell the IR  renormalons  ambiguities in $D_{PT}$) additionnal {\em unambiguous}
contributions originating {\em  from the UV region }, hence unrelated  to IR renormalons and  the
OPE , e.g.:
\begin{eqnarray*}\ \ \ \ \ \ \ \ \ \ \ \ \ \ \int_{Q^2}^\infty{dk^2\over k^2}\
\delta\alpha_s^{PT}(k^2)\ \varphi\left(k^2\over Q^2\right)&\simeq&-\sum_{p=1}^\infty  A_p
b_p^{PT} \left(-\ {\Lambda^2\over Q^2}\right)^p\\  Q^2&\gg&\Lambda^2\ \ \ \ \ \ \ \ \ \ \ \ \ \ \
\ \ \ \ \ \ \ \ \ \ \ \ \ \ \ \ \ \ \ \ \ \ 
\ \ \ \ \ \ (6.6)
\end{eqnarray*} where:
$$A_p = \int_{Q^2}^\infty{dk^2\over k^2}\ \varphi\left(k^2\over Q^2\right)\ \left(Q^2\over
k^2\right)^p\eqno(6.7)$$ is a number, and I used the asymptotic expansion eq.(3.13) of
$\delta\alpha_s^{PT}(k^2)$ (throughout this section, I also assume $\alpha_s^{PT}$ has no
renormalons, so that the
$b_i^{PT}$'s themselves are real and unambiguous).

\noindent Let us derive the result for $\delta D_{PT}$ in the typical (cf. eq.(5.1)) case where :
$$D(Q^2)=\int_{0}^{Q^2}n{dk^2\over k^2}\ \alpha_s(k^2)\ \left(k^2\over Q^2\right)^n\eqno(6.8)$$ 

\noindent i)\underline {Assume first $n$ is an integer}. One proceeds by disentangling [30-32]
long from short distances and split [18] the integral in eq.(6.8) at the arbitrary IR scale
$\Lambda_I^2=c\Lambda^2$ \\$(c > 1)$ :
\begin{eqnarray*}\ \ \ \ \delta D_{PT}(Q^2)& = &\int_0^{\Lambda_I^2}n{dk^2\over k^2}\
\delta\alpha_s^{PT}(k^2)\ \left(k^2\over Q^2\right)^n + \int_{\Lambda_I^2}^{Q^2}n{dk^2\over k^2}\
\delta\alpha_s^{PT}(k^2)\ \left(k^2\over Q^2\right)^n\\  &\equiv& \delta D_{ld}^{PT} + \delta
D_{sd}^{PT}\ \ \ \ \ \ \ \ \ \ \ \ \ \ \ \ \ \ \ \ \ \ \ \ \
\ \ \ \ \ \ \ \ \ \ \ \ 
\ \ \ \ \ \ \ \ \ \ \ \ \ \ \ \ \ \ \ \ \ \ (6.9)\end{eqnarray*} The ``long distance'' part
$\delta D_{ld}^{PT}$ contributes the (ambiguous) power correction:
$$\delta D_{ld}^{PT}(Q^2)= K_n^{PT}(\Lambda_I^2) \left(\Lambda^2\over Q^2\right)^n\eqno(6.10)$$ 
with
$$ K_n^{PT}(\Lambda_I^2)=\int_0^{\Lambda_I^2}n{dk^2\over k^2}\ \left(k^2\over \Lambda^2\right)^n\
\delta\alpha_s^{PT}(k^2)\eqno(6.11)$$ This ``IR'' power correction cancells the $z_n=n/\beta_0$
IR  renormalon ambiguity present in
$D_{PT}$, and is best combined with the similar contribution to $D_{PT}$ from the same
integration range to yield the unambiguous ${\cal O}\left((\Lambda^2/Q^2)^n\right)$ power
correction:
$$\int_0^{\Lambda_I^2}n{dk^2\over k^2}\ \alpha_{s,reg}^{PT}(k^2)\ \left(k^2\over Q^2\right)^n=
\left(\Lambda^2\over Q^2\right)^n \left[\int_0^{\Lambda_I^2}n{dk^2\over k^2}\
\alpha_{s,reg}^{PT}(k^2)\
\left(k^2\over \Lambda^2\right)^n\right]\eqno(6.12)$$  On the other hand, the ``short distance''
part yields:
$$\delta D_{sd}^{PT}(Q^2)=-\sum_{p\neq n}{n b_p^{PT}\over n-p} \left(-\ {\Lambda^2\over
Q^2}\right)^p - \left(n b_n^{PT} \ln {Q^2\over
\Lambda^2} + const\right) \left(-\ {\Lambda^2\over Q^2}\right)^n\eqno(6.13)$$ Eq.(6.13) may be
easily obtained by substituting eq.(3.13) into the second integral in eq.(6.9) . I give a more
general derivation, where it is not necessary to assume that the expansion eq.(3.13) is valid
down to $k^2=\Lambda_I^2$. It is convenient to separate the first n terms of the asymptotic
expansion and define:
$$\delta\alpha_s^{PT}(k^2)\equiv -\sum_{p=1}^n b_p^{PT} \left(-{\Lambda^2\over k^2}\right)^p +
\left[\delta\alpha_s^{PT}(k^2)\right]_{(n)}\eqno(6.14)$$  Then:
$$\delta D_{sd}^{PT}(Q^2)=-\sum_{p=1}^n b_p^{PT} \left(-\ {\Lambda^2\over Q^2}\right)^p
\int_{\Lambda_I^2}^{Q^2} n{dk^2\over k^2}\left(k^2\over Q^2\right)^{n-p} +
\int_{\Lambda_I^2}^{Q^2} n{dk^2\over k^2}\ \left[\delta\alpha_s^{PT}(k^2)\right]_{(n)}\
\left(k^2\over Q^2\right)^n\eqno(6.15)$$ The second integral in eq.(6.15) is now UV convergent,
and can be expressed as:
\begin{eqnarray*}\ \ \ \ \ \ \ \ \int_{\Lambda_I^2}^{Q^2} n{dk^2\over k^2}\
\left[\delta\alpha_s^{PT}(k^2)\right]_{(n)}\ \left(k^2\over Q^2\right)^n &=&
\int_{\Lambda_I^2}^\infty n{dk^2\over k^2}\ \left[\delta\alpha_s^{PT}(k^2)\right]_{(n)}\
\left(k^2\over Q^2\right)^n \\ & - &\int_{Q^2}^{\infty} n{dk^2\over k^2}\
\left[\delta\alpha_s^{PT}(k^2)\right]_{(n)}\ \left(k^2\over Q^2\right)^n
\ \ \ (6.16)\end{eqnarray*}  But:
\begin{eqnarray*}\ \ \ \ \ \ \ \ \ \ \ \ \ \ \ \ \ \ \int_{\Lambda_I^2}^\infty n{dk^2\over k^2}\
\left[\delta\alpha_s^{PT}(k^2)\right]_{(n)}\ \left(k^2\over Q^2\right)^n & = &
\left(\Lambda^2\over Q^2\right)^n \int_{\Lambda_I^2}^\infty n{dk^2\over k^2}\
\left[\delta\alpha_s^{PT}(k^2)\right]_{(n)}\ \left(k^2\over \Lambda^2\right)^n \\ &=& const
\left(\Lambda^2\over Q^2\right)^n\ \ \ \ \ \ \ \ \ \ \ \ \ \ \ \ \ \  (6.17)\end{eqnarray*}  
whereas, using eq.(3.13):
$$\int_{Q^2}^{\infty} n{dk^2\over k^2}\ \left[\delta\alpha_s^{PT}(k^2)\right]_{(n)}\
\left(k^2\over Q^2\right)^n = \sum_{p > n} {nb_p^{PT}\over n-p} \left(-\ {\Lambda^2\over
Q^2}\right)^p\eqno(6.18)$$ On the other hand, the first term in eq.(6.15) gives:
\begin{eqnarray*}-\sum_{p=1}^n b_p^{PT} \left(-{\Lambda^2\over Q^2}\right)^p
\int_{\Lambda_I^2}^{Q^2} n{dk^2\over k^2}\left(k^2\over Q^2\right)^{n-p}&=&-\sum_{p <
n}{nb_p^{PT}\over n-p} \left(-{\Lambda^2\over Q^2}\right)^p\\ &-&\left(n b_n^{PT} \ln {Q^2\over
\Lambda^2}+const\right)\left(-{\Lambda^2\over Q^2}\right)^n\ \ \ (6.19)\end{eqnarray*}
Eq.[(6.15)-(6.19)] then yields eq.(6.13). Combining eq.(6.10) and (6.13), one ends up with:
$$\delta D_{PT}(Q^2)=-\sum_{p\neq n}{nb_p^{PT}\over n-p} \left(-\ {\Lambda^2\over Q^2}\right)^p -
n \left(b_n^{PT} \ln {Q^2\over
\Lambda^2} +  D_n^{PT}\right) \left(-\ {\Lambda^2\over Q^2}\right)^n\eqno(6.20)$$ where the
constant $ D_n^{PT}$ is independant of the arbitrary IR scale $\Lambda_I$, but complex and
ambiguous - thus cancelling the IR renormalon ambiguity arising from a simple pole at
$z=z_n$ in $\tilde{D}(z)$  , and cannot be computed straightforwardly (apart from its imaginary
part, see Appendix A) from the asymptotic expansion eq.(3.13).

\noindent Furthermore, $\delta D_{PT}(Q^2)$ for a general observable as in eq.(2.1) may be 
easily obtained from eq.(6.6) and (6.20) , splitting the integral in eq.(6.1) at $k^2=Q^2$, and
expanding the kernel for
$k^2\leq Q^2$ :
$$\varphi(k^2/Q^2) = \sum_{n=1}^\infty n c_n \left({k^2\over Q^2}\right)^n\eqno(6.21)$$ (where I
assumed for simplicity absence of  logarithmic terms, see point ii) below). One gets:
$$\delta D_{PT}(Q^2)=-\sum_n b_n^{PT} d_n \left(-{\Lambda^2\over Q^2}\right)^n -\sum_n n c_n
\left( b_n^{PT}
 \ln {Q^2\over\Lambda^2} + D_n^{PT}\right)\left(-{\Lambda^2\over Q^2}\right)^n \eqno(6.22)$$
where 
$$d_n=A_n + \sum_{p\neq n}{p c_p\over p-n}\eqno(6.23)$$ depends only on the $\varphi$ kernel.

\noindent ii) I also quote the analoguous result for the log-enhanced kernel where $n$ is
integer, but:
$$D(Q^2)=\int_{0}^{Q^2}n{dk^2\over k^2}\ \alpha_s(k^2)\ \left(k^2\over Q^2\right)^n\ \ln{Q^2\over
k^2}\eqno(6.24)$$ With similar methods, one gets:
$$\delta D_{PT}(Q^2)=-\sum_{p\neq n}{n b_p^{PT}\over (n-p)^2} \left(-\ {\Lambda^2\over
Q^2}\right)^p - n\left({1\over 2}b_n^{PT} \ln^2 {Q^2\over
\Lambda^2} +  E_n^{PT}\ln{Q^2\over
\Lambda^2} + F_n^{PT}\right) \left(-\ {\Lambda^2\over Q^2}\right)^n\eqno(6.25)$$ where $E_n^{PT}$
and $F_n^{PT}$ are complex and ambiguous (they cancell the IR renormalon ambiguity arising from a
double pole in $\tilde{D}(z)$).

\noindent iii)\underline {Non-integer $n$}: in such a case it is no more necessary to introduce
an intermediate scale $\Lambda_I$, and log-enhanced terms are absent. Specifically, suppose
$0<n<1$. Then, with $D(Q^2)$ as in eq.(6.8):
\begin{eqnarray*}\ \ \ \ \ \ \ \ \delta D_{PT}(Q^2) &=&
\int_0^\infty n{dk^2\over k^2}\ \delta\alpha_s^{PT}(k^2)\ \left(k^2\over Q^2\right)^n -
\int_{Q^2}^{\infty} n{dk^2\over k^2}\ \delta\alpha_s^{PT}(k^2)\ \left(k^2\over Q^2\right)^n \\
&\equiv &  K_n^{PT} \left({\Lambda^2\over Q^2}\right)^n - \sum_{p =1}^{\infty} {nb_p^{PT}\over
n-p} \left(-\ {\Lambda^2\over Q^2}\right)^p
\ \ \ \ \ \ \ \ \ \ \ \ \ \ \ \ \ \ \ \ \ \ \ \ \ \ \ (6.26)\end{eqnarray*} where $ K_n^{PT}$ is
again a complex, ambiguous constant corresponding to a simple IR renormalon pole at $z=z_n$ :
$$ K_n^{PT} \equiv \int_0^\infty n{dk^2\over k^2}\left({k^2\over\Lambda^2}\right)^n
\delta\alpha_s^{PT}(k^2)\equiv  K_n^{PT}(\Lambda_I^2=\infty)\eqno(6.27)$$

As an application of the above results, one can derive the ``perturbative'' power corrections
generated by the regularized one loop coupling eq.(3.1). Using:
$$\delta\alpha_s^{PT}(k^2)_{\mid one-loop}=-{1\over\beta_0}{{\Lambda^2\over
k^2}\over1-{\Lambda^2\over k^2}}\eqno(6.28)$$ one finds they are given by the Beneke-Braun
formula [6]:
\begin{eqnarray*}\ \ \ \ \ \ \ \delta D_{PT}(Q^2)_{\mid one-loop}& =& -{1\over\beta_0}
\int_{0}^\infty{dk^2\over k^2}\ {{\Lambda^2\over k^2}\over1-{\Lambda^2\over k^2}}\
\varphi\left(k^2\over Q^2\right) \\ &\equiv&-{1\over\beta_0} \left[{\cal F}(-{\Lambda^2\over
Q^2})-{\cal F}(0)\right]\ \ \ \ \ \ \ \ \ \ \ \ \ \ \ \ \ \ \
\ \ \ \ \ \ \ \ \ \ \ \ \ \ \ (6.29)\end{eqnarray*} since eq.(2.6) implies:
$${\cal F}({\mu^2\over Q^2})-{\cal F}(0) = \int_{0}^\infty{dk^2\over k^2}\ {-{\mu^2\over
k^2}\over1+{\mu^2\over k^2}}\ \varphi\left(k^2\over Q^2\right)\eqno(6.30)$$ i.e. the once
subtracted dispersion relation eq.(6.30) for the characteristic function  may be seen as a
peculiar case of the formula eq.(6.4) for $\delta D_{PT}(Q^2)$, with the substitutions: $\Lambda^2
\rightarrow -\mu^2 $ and
$\delta\alpha_s^{PT}(k^2)\rightarrow {-{\mu^2\over k^2}\over1+{\mu^2\over k^2}}$  (which imply
$b_p^{PT} \rightarrow (-1)^{p+1}$ in eq.(3.13)). If $\varphi$ behaves as in eq.(6.21), this
observation implies (eq.(6.22)) the small $\mu^2$ behavior:
$${\cal F}({\mu^2\over Q^2})-{\cal F}(0)= \sum_n d_n\left(-\ {\mu^2\over Q^2}\right)^n +\sum_n n 
c_n
\left(\ln {Q^2\over\mu^2}+{\tilde d}_n \right)\left(-\ {\mu^2\over Q^2}\right)^n\eqno(6.31)$$
where the constant ${\tilde d}_n$ can be computed explicitly. If $0<n<1$, one gets instead from
eq.(6.26) (for $D(Q^2)$ as in eq.(6.8)):
$${\cal F}({\mu^2\over Q^2})-{\cal F}(0)=-\ {\pi n\over\sin(\pi n)} \left({\mu^2\over
Q^2}\right)^n + 
\sum_{p =1}^{\infty} {n\over n-p} \left(-\ {\mu^2\over Q^2}\right)^p\eqno(6.32)$$ where I used
the identity:
$$\int_0^\infty n{dk^2\over k^2}\left({k^2\over\mu^2}\right)^n {\mu^2\over \mu^2+k^2}\equiv {\pi
n\over\sin(\pi n)}\eqno(6.33)$$

\noindent It is interesting to note that any ${\cal O}\left((k^2/Q^2)^n\right)$ term in the low
energy expansion of the kernel $\varphi\left(k^2\over Q^2\right)$ is in one-to-one
correspondance, for $n$ integer, with a non-analytic term $
\left(\mu^2\over Q^2\right)^n \ln {Q^2\over \mu^2}$ in ${\cal F}({\mu^2\over Q^2})$ , which
explains the connection [4,6]  between non-analytic terms in the characteristic function and IR
renormalons. However these non-analytic terms are not quite of IR origin, since they arise from
the analogue in eq.(6.30) of the
$\delta D_{sd}^{PT}$ piece of eq.(6.9), and {\em not} from (the analogue of) $\delta
D_{ld}^{PT}$: they  correspond to an {\em UV} enhancement (letting $Q^2\rightarrow \infty$ in
eq.(6.9)) rather than to an IR one. More generally, the {\em leading} log terms in ${\cal
F}({\mu^2\over Q^2})$ and $\delta D_{PT}(Q^2)$, and in particular the {\em analytic} terms if
there are no log (which implies, barring cases where $b_n^{PT}=0$ (see below), the vanishing of
the corresponding coefficient $c_n$) are  unambiguous and of short distance origin (see 
eq.(6.22), (6.25) and (6.31)). Comparing eq.(6.22) and (6.31) show they are simply related  by a 
$b_n^{PT}$ factor.   On the other hand, the {\em sub-leading} log terms (in particular the
constant terms associated to a log) are ambiguous and of (partially)  IR origin. It is actually
not possible,  without reintroducing an arbitray IR cut-off $\Lambda_I$, to disentangle 
unambiguously terms of IR and UV origin within the sub-leading log terms of eq.(6.22) and (6.31) 
(the exception to the previous statement is the case $n\neq$ integer, where no IR cut-off
$\Lambda_I$ needs to be introduced, see eq.(6.26)). Thus, for $n$ integer,   non-analytic terms
are ``related to'', but
 do not really arise from, long distances (this is also apparent from the fact that their
coefficient is proportionnal (eq.(6.22)) to the product $c_n b_n^{PT}$ of a long distance $\times$
a  short distance parameter).

\noindent The previous remarks suggest a simple generalization of eq.(6.29) to an arbitrary
coupling: assume the {\em leading} term is  in eq.(6.22) is an
${\cal O}\left(({\Lambda^2/Q^2})^n\right)$   power correction  {\em entirely} of short distance
origin,  i.e. that $c_i=0$ for $i<n$ and $\varphi$ is ${\cal O}\left((k^2/Q^2)^n\right)$ at small
$k^2$. Then :
\begin{eqnarray*}\ \ \ \ \ \ \ \ \ \ \ \ \ \ \delta D_{PT}(Q^2)&\simeq&- b_n^{PT} d_n \left(-\
{\Lambda^2\over Q^2}\right)^n\\  Q^2&\gg&\Lambda^2\ \ \ \ \ \ \ \ \ \ \ \ \ \ \ \ \ \ \ \ \ \ \ \
\ \ \ \ \ \ \ \ \ \ \ \ \ 
\ \ \ \ \ \ \ \ \ \ \ \ \ \ \ \ \ \ \ \ \ \ \ \ \ (6.34)
\end{eqnarray*} while eq.(6.31) shows  that the leading small $\mu^2$ behavior of
${\cal F}({\mu^2\over Q^2})$ is analytic and given by :
\begin{eqnarray*}\ \ \ \ \ \ \ \ \ \ \ \ \ \ {\cal F}({\mu^2\over Q^2})-{\cal
F}(0)&\simeq&d_n\left(-\ {\mu^2\over Q^2}\right)^n\\ 
\mu^2&\ll&Q^2\ \ \ \ \ \ \ \ \ \ \ \ \ \ \ \ \ \ \ \ \ \ \ \ \ \ \ \ \ \ \ \ \ \ \ \ \ 
\ \ \ \ \ \ \ \ \ \ \ \ \ \ \ \ \ (6.35)
\end{eqnarray*} Eq.(6.34)-(6.35)  agree with eq.(6.29) in the one loop case, where (eq.(4.25))\\ 
$b_n^{PT}=(-1)^n{1\over
\beta_0}$~.    

In the complementary case where a peculiar $b_n^{PT}$ vanishes while $c_n\neq 0$, it is not
necessary any more to introduce an IR cut-off
$\Lambda_I$, since all integrals in eq.(6.15) are separately IR convergent. Setting $\Lambda_I=0$ 
( with
$D(Q^2)$ as in eq.(6.8)) one gets:
$$\delta D_{PT}(Q^2)=-\sum_{p\neq n}{nb_p^{PT}\over n-p} \left(-\ {\Lambda^2\over Q^2}\right)^p +
\int_0^\infty n{dk^2\over k^2}\ \left[\delta\alpha_s^{PT}(k^2)\right]_{(n)}\ \left(k^2\over
Q^2\right)^n \eqno(6.36)$$ where the last integral is an IR ${\cal O}\left((1/Q^2)^n\right)$
power correction (with all other contributions arising from short distances). Note that a
non-analytic term is still present in ${\cal F}({\mu^2\over Q^2})$, and is actually crucial to
reproduce the IR power correction in $\delta D_{PT}(Q^2)$ (see section 7, where a more general
derivation of these results, which relies directly on the representation eq.(2.17) and does not
assume the dispersion relation eq.(6.30), is given).

\subsection{Non-perturbative power corrections} In [7], a condition of sufficiently fast  UV
damping (i.e. of an exponential  or at least of a rather high power suppression at large $k^2$)
was imposed on the  ``non-perturbative'' modification
$\delta\alpha_s^{NP}(k^2)$ . The assumption that $\delta\alpha_s^{NP}(k^2)$ is essentially
restricted to low $k^2$ was motivated by the  ideology of ``soft confinement'' of [20], who put
forward the idea of gluon condensation as an essentially IR phenomenon, which could be described
entirely within the OPE . However there is no fundamental reason, as is by now widely
appreciated,  that all power contributions should be of IR origin, and in fact the dispersive
framework strongly suggests the existence of power contributions arising from short distances, as
examplified in section 6.1 through the simplest dispersive regularization procedure. For the sake
of generality, I shall therefore relax this assumption . The split eq.(2.14) of
$\delta\alpha_s$ into ``perturbative'' and ``non-perturbative'' components now becomes a matter of
convention, since one can no more argue as in section 2 that
$\delta\alpha_s\simeq\delta\alpha_s^{PT}$ at large $k^2$. I however still  assume  that
$\delta\alpha_{eff}^{NP}(\mu^2)$ itself is  exponentially suppressed at large $\mu^2$, in order
to be able to expand under the integral in eq.(2.9): this is a rather strong restriction, but
represents a straightforward extension of the framework of [7]. Then eq.(2.9) yields the
asymptotic expansion  :
$$\delta\alpha_s^{NP}(k^2)=-\sum_{n=1}^\infty b_n^{NP} \left(-\ {\Lambda^2\over
k^2}\right)^n\eqno(6.37) $$ where the integer moments:
$$b_n^{NP}=\int_0^\infty\ n{d\mu^2\over\mu^2}\ \left({\mu^2\over \Lambda^2}\right)^n\
\delta\alpha_{eff}^{NP} (\mu^2)\eqno(6.38)$$  are no more required to vanish as in [7]. It
follows that in eq.(2.14):
$$\delta\alpha_s(k^2)=-\sum_{n=1}^\infty b_n \left(-\ {\Lambda^2\over k^2}\right)^n\eqno(6.39) $$ 
with:
$$b_n\equiv b_n^{PT}+b_n^{NP}\eqno(6.40) $$ It is clear that all results of section 6.1 also
apply to $\delta D_{NP}(Q^2)$ (eq.(2.10a)) or to the {\em total} power contribution:
$$\delta D(Q^2)\equiv \int_{0}^\infty{dk^2\over k^2}\ \delta\alpha_s(k^2)\ \varphi\left(k^2\over
Q^2\right)\eqno(6.41)$$ if one substitutes
$\delta\alpha_s^{PT}(k^2)$ with $\delta\alpha_s^{NP}(k^2)$ or $\delta\alpha_s(k^2)$, and
$b_n^{PT}$ with $b_n^{NP}$ or $b_n$ (of course for $\delta\alpha_s^{NP}$  all subleading logs
constants are also unambiguous). In particular, the conclusion of section 6.1 that it is not
possible in general to disentangle in an unambiguous way for $n$ integer the power corrections of
IR origin from those which arise from short distances is also valid here. 

The present general framework is still compatible with the  assumption [7,20] that
$\delta\alpha_s^{NP}(k^2)$ (or even the {\em total} $\delta\alpha_s(k^2)$)  is restricted to low
$k^2$, but this question has to be decided by fitting the data, rather then imposed a priori .
For instance, if one assumes that
$\delta\alpha_s^{NP}(k^2)\equiv\left[\delta\alpha_s^{NP}(k^2)\right]_{(n)}$, i.e. that
$b_p^{NP}=0$ for $1\leq p\leq n$, then one obtains immediately for the observable of eq.(6.8)
(from the analogue of eq.(6.36), see also eq.(6.26)) :
\begin{eqnarray*}\ \ \ \ \ \ \ \ \delta D_{NP}(Q^2) &=&
\int_0^\infty n{dk^2\over k^2}\ \delta\alpha_s^{NP}(k^2)\ \left(k^2\over Q^2\right)^n -
\int_{Q^2}^{\infty} n{dk^2\over k^2}\ \delta\alpha_s^{NP}(k^2)\ \left(k^2\over Q^2\right)^n \\
&\equiv &  K_n^{NP} \left({\Lambda^2\over Q^2}\right)^n - \sum_{p>n} {nb_p^{NP}\over n-p}
\left(-\ {\Lambda^2\over Q^2}\right)^p
\ \ \ \ \ \ \ \ \ \ \ \ \ \ \ \ \ \ \ \ \ \ \ \ \ \ \ (6.42)\end{eqnarray*} where:
$$ K_n^{NP} \equiv \int_0^\infty n{dk^2\over k^2}\left({k^2\over\Lambda^2}\right)^n
\delta\alpha_s^{NP}(k^2)\eqno(6.43)$$ i.e. in this example the leading power correction is indeed
of IR origin and controlled by the OPE . Imposing similarly that $\delta\alpha_s^{NP}(k^2)$ be
exponentially suppressed , i.e. that $b_n^{NP}=0$ for all $n$'s, leads to the  large $Q^2$ result
(for the general observable of eq.(6.1), assuming eq.(6.21)):
$$\delta D_{NP}(Q^2) = \sum_{n=1}^\infty c_n K_n^{NP} \left({\Lambda^2\over
Q^2}\right)^n\eqno(6.44)$$ All  non-perturbative power corrections, arising essentially from the
infrared, are then in one to one correspondence with a term $c_n$ in the low energy expansion of
the Feynman diagram kernel, hence [30-32] with a related operator in the OPE . Alternatively, one
could impose that the {\em total} $\delta\alpha_s(k^2)$ be exponentially suppressed, by requiring
that $b_n^{NP}=-b_n^{PT}$ for all $n$'s (given the $b_n^{PT}$'s,  a theorem [33] guarantees there
is an {\em infinity} of solutions $\delta\alpha_{eff}^{NP} (\mu^2)$ for the resulting moment
problem following from eq.(6.38); for instance (see [34] for a related suggestion), these
constraints may be fulfilled by expressing $\delta\alpha_s^{NP}(k^2)$  as an (eventually
infinite) sum of time-like  poles). Nevertheless,  all such restrictions have no fundamental 
basis, barring the (arbitrary) requirement that only those power corrections to the Borel sum
which are controlled by the OPE should be present.

\section{Minkowskian observables}  For Euclidean observables, the alternative representation in
term of the characteristic function (eq.(2.5 b), (2.10 b) and (2.17)), although technically
convenient, is not really indispensable. The situation is different for  Minkowskian observables
$R(Q^2)$ (such as cross sections or inclusive decay rates), for which the representation eq.(2.1)
is not in general available. In such cases the characteristic function
${\cal F}_R({\mu^2\over Q^2})$ is usually given by two distinct pieces, for instance:
$${\cal F}_R({\mu^2\over Q^2})=\left\{\begin{array}{ll}{\cal F}_{(-)}({\mu^2\over Q^2})&\mu^2<
Q^2\\ {\cal F}_{(+)}({\mu^2\over Q^2})&\mu^2> Q^2\end{array}\right.\eqno(7.1)$$  (where ${\cal
F}_{(-)}$ is the sum of a real and a virtual contribution, while ${\cal F}_{(+)}$ contains only
the virtual contribution), and thus cannot satisfy the dispersion relation eq.(2.6). Then $R$
and  $\delta R_{NP}$ are given directly by eq.(2.5 b) (respectively (2.10 b)) (with
$D\rightarrow R$ and $\dot{{\cal F}}\rightarrow \dot{{\cal F}_R}$) , i.e.:
$$R(Q^2)=\int_0^\infty{d\mu^2\over\mu^2}\ \alpha_{eff}(\mu^2)\ \dot{{\cal F}_R}({\mu^2\over
Q^2})\eqno(7.2)$$ and similarly:
$$R_{reg}^{PT}(Q^2)=\int_0^\infty{d\mu^2\over\mu^2}\ \alpha_{eff}^{PT}(\mu^2)\ \dot{{\cal
F}_R}({\mu^2\over Q^2})\eqno(7.3)$$ where 
$\alpha_{eff}^{PT}$ is obtained from the discontinuity of the (Borel summed) 
$\alpha_s^{PT}$  as explained in section 4 (I assume $\tilde{\alpha}_s(z)$ has no renormalons),
and is IR finite . 

Let us derive eq.(7.2)-(7.3) in the peculiar case where $R(Q^2)$ is related to the time-like
discontinuity of an Euclidean observable $D(Q^2)$  which satisfies the dispersion relation~:
$$D(Q^2)=Q^2\int_0^\infty{dQ'^2\over(Q'^2+Q^2)^2}\ R(Q'^2)\eqno(7.4)$$ Then:
$${dR\over d\ln Q^2}=\rho_D(Q^2)\eqno(7.5)$$ where $\rho_D(Q^2)\equiv  -\frac{1} {2\pi i}
Disc\{D(-Q^2)\}$ ($Q^2>0$) is the time-like ``spectral density'' of $D(Q^2)$. On the other hand
,  if $D(Q^2)$   satisfies the representation eq.(2.1) , the corresponding $\rho_D(Q^2)$ is given
in term of the spectral density $\rho(\mu^2)$ of $\alpha_s(k^2)$ by:
$$\rho_D(Q^2)=\int_{0}^\infty{d\mu^2\over
\mu^2}\ \rho(\mu^2)\ \varphi\left(\mu^2\over Q^2\right)\eqno(7.6)$$ Eq.(7.6) follows [18]  by
performing the change of variable  $x=k^2/Q^2$ in eq.(2.1), and performing the analytic
continuation to the time-like region with the new integrand:
$$D(Q^2)=\int_{0}^\infty{dx\over x}\ \alpha_s(x Q^2)\ \varphi(x)\eqno(7.7)$$ The corresponding
$R(Q^2)$ from eq.(7.5) is then given by [18]:
$$R(Q^2)=\int_{0}^\infty{d\mu^2\over
\mu^2}\ \alpha_{eff}(\mu^2)\ \varphi\left(\mu^2\over Q^2\right)\eqno(7.8)$$ A similar argument 
applied to $D_{reg}^{PT}(Q^2)$ (eq.(2.12))  yields (since
$Disc\{\alpha_{s,reg}^{PT}(-\mu^2)\}\equiv Disc\{\alpha_s^{PT}(-\mu^2)\}$):
$$\rho_{D,reg}^{PT}(Q^2)=\int_{0}^\infty{d\mu^2\over
\mu^2}\ \rho_{PT}(\mu^2)\ \varphi\left(\mu^2\over Q^2\right)\eqno(7.9)$$ and:
$$R_{reg}^{PT}(Q^2)=\int_{0}^\infty{d\mu^2\over
\mu^2}\ \alpha_{eff}^{PT}(\mu^2)\ \varphi\left(\mu^2\over Q^2\right)\eqno(7.10)$$ where
$\rho_{D,reg}^{PT}(Q^2)\equiv~-\frac{1} {2\pi i} Disc\{{D_{reg}^{PT}}(-Q^2)\}$ and:
$${dR_{reg}^{PT}\over d\ln Q^2}=\rho_{D,reg}^{PT}(Q^2)\eqno(7.11)$$ Eq.(7.8) and (7.10) suggest
that [18]:
$$\dot{{\cal F}_R}({\mu^2\over Q^2})\equiv\varphi\left(\mu^2\over Q^2\right)\eqno(7.12)$$ I
complete the argument and show that~: i) $R_{reg}^{PT}(Q^2)$ in eq.(7.10) has the correct 
perturbative expansion and Borel transform expected from eq.(7.4)  and ii) $D_{reg}^{PT}(Q^2)$ in
eq.(2.12) or (2.17) is related to
$R_{reg}^{PT}(Q^2)$  by the dispersion relation eq.(7.4). 

\noindent i) Eq.(7.4) implies [27] the relation between the Borel transforms of absorptive and
dispersive parts:
$$\tilde{R}(z)= {\sin(\pi\beta_0 z)\over\pi\beta_0 z }\ \tilde{D}(z)\eqno(7.13)$$ This relation
is indeed satisfied since the Borel sum corresponding to  eq.(7.10) is :
$$R_{PT}(Q^2)=\int_0^\infty dz \exp\left(-z\beta_0\ln{Q^2 \over
\Lambda^2}\right)\ \tilde{R}(z)\eqno(7.14)$$ with (an analogue of eq.(6.5b)):
$$\tilde{R}(z)= \tilde{\alpha}_{eff}(z)\  \int_0^\infty{d\mu^2\over \mu^2}\ \varphi
\left(\mu^2\over Q^2\right)\ \exp\left(-z\beta_0\ln{\mu^2\over Q^2 }\right)\eqno(7.15)$$
Eq.(7.14)-(7.15) are obtained [23] by freely substituting $\alpha_{eff}^{PT}$ by its Borel
representation eq.(4.4) into eq.(7.10), and  permutting  the orders of integration. In this case
however, because $\alpha_{eff}^{PT}$ has a non-trivial IR fixed point,
$R_{reg}^{PT}$ differs [23,24], as we shall see below, from its Borel sum $R_{PT}$ (in sharp
contrast with $D_{PT}$ in eq.(6.3)). Eq.(7.15), together with eq.(4.6) and (6.5b),  reproduce
eq.(7.13). 

\noindent ii) Substituting $R$ in the dispersion relation eq.(7.4) with $R_{reg}^{PT}$
(eq.(7.10)), one finds, after permutting the orders of integration:
\begin{eqnarray*}\ \ \ \ \ \ \ \ \ \ \ \ \ \ \ \ \ D_{reg}^{PT}(Q^2)&\equiv
&Q^2\int_0^\infty{dQ'^2\over(Q'^2+Q^2)^2}\ R_{reg}^{PT}(Q'^2)\\ 
&=&\int_0^\infty{d\mu^2\over\mu^2}\ \alpha_{eff}^{PT}(\mu^2)\ \dot{{\cal F}}({\mu^2\over Q^2})\
\ \ \ \ \ \ \ \ \ \ \ \ \ \ \ \ \ \ \ \ \ \ \ \ \ \ \
\ \ \ \ (7.16)\end{eqnarray*}  with:
\begin{eqnarray*}\ \ \ \ \ \ \ \ \ \
\ \ \ \ \ \
\ \ \ \ \dot{{\cal F}}({\mu^2\over Q^2})&\equiv &Q^2\int_0^\infty{dQ'^2\over(Q'^2+Q^2)^2}\
\varphi\left(\mu^2\over Q'^2\right)\\  &=&\mu^2\int_0^\infty{dk^2\over(k^2+\mu^2)^2}\
\varphi\left(k^2\over Q^2
\right)\
\ \ \ \ \ \ \ \ \ \ \ \ \ \ \ \ \ \ \ \ \ \ \ \ \ \ \
\ (7.17)\end{eqnarray*}  where the change of variable ${\mu^2\over Q'^2}={k^2\over Q^2}$ has been
performed in the second step. Eq.(7.17) indeed agrees with the expected relation for $\dot{{\cal
F}}$ obtained by taking the
$\mu^2$-derivative of eq.(2.6), and shows that eq.(7.16) reproduces eq.(2.17) , hence eq.(2.12)~.

Finally, let us justify eq.(A.1) of Appendix A, i.e. the statement that:
$$Disc\{D_{reg}^{PT}(-Q^2)\} = Disc\{D_{PT}(-Q^2)\} \ \ \ \ \ (Q^2>0)$$ This result follows
immediately by applying  to $D_{PT}(Q^2)$ in eq.(6.3)  the same argument [18] which leads to
eq.(7.9), and reflects the basic feature of the dispersive regularization procedure that 
$Disc\{\alpha_{s,reg}^{PT}(-\mu^2)\}\equiv Disc\{\alpha_s^{PT}(-\mu^2)\}$. However now there is a
caveat, since $\alpha_s^{PT}(k^2)$ does not satisfy the dispersion relation eq.(2.2). In
particular, it could have {\em complex} singularities in the 
$k^2$ plane (in addition to the standard space-like Landau singularity ), which would make
eq.(7.7) (with $\alpha_s\rightarrow \alpha_s^{PT}$) meaningless at complex $Q^2$, and obstruct
the analytic continuation to the time-like region. The procedure of [18] seems however to be 
safe if one {\em assumes} the absence of such singularities (as is the case for the one-loop
coupling ).

\subsection{Perturbative power corrections} From the results of section 6, we know that
$D_{reg}^{PT}$  (eq.(2.17)) differs from its Borel sum $D_{PT}$ . It is therefore natural to
expect that also here 
$R_{reg}^{PT}$ differs from the corresponding Borel sum $R_{PT}$ (eq.7.14) by ``perturbative''
power corrections $\delta R_{PT}$. To determine them, one can proceed as in section 4 and
Appendix A, and split the integral in eq.(7.3) at $\mu^2=\Lambda^2$ :
\begin{eqnarray*}\ \ \ \ \ \ \ \ \ R_{reg}^{PT}(Q^2) &=&
\int_0^{\Lambda^2}{d\mu^2\over\mu^2}\ \alpha_{eff}^{PT}(\mu^2)\ \dot{{\cal F}_R}({\mu^2\over
Q^2})+\int_{\Lambda^2}^\infty{d\mu^2\over\mu^2}\ \alpha_{eff}^{PT}(\mu^2)\
\dot{{\cal F}_R}({\mu^2\over Q^2})\\ &\equiv&R_{reg,<}^{PT}(Q^2)+R_>^{PT}(Q^2)\ \ \ \ \ \ \ \ \ \
\ \ \ \ \ \ \ \ \
\ \ \ \ \ \ \ \ \ \ \ \ \ \ \ \ \ \ \ \ \ \ \ \ \ (7.18)\end{eqnarray*}   Using  the Borel
representation of $\alpha_{eff}^{PT}$ (eq.(4.4)) , $R_>^{PT}$ can be written as:
$$R_>^{PT}(Q^2)=R_{PT}(Q^2) - R_<^{PT}(Q^2)\eqno(7.19)$$ where (for $Q^2>\Lambda^2$):
$$R_<^{PT}(Q^2)\equiv\int_0^\infty dz\ \tilde{\alpha}_{eff}(z)\ 
\left[\int_0^{\Lambda^2}{d\mu^2\over \mu^2}\
\dot{{\cal F}_R}({\mu^2\over Q^2})\ \exp\left(-z\beta_0\ln{\mu^2\over
\Lambda^2}\right)\right]\eqno(7.20)$$ Thus:
$$R_{reg}^{PT}(Q^2)=R_{PT}(Q^2)+\delta R_{PT}(Q^2)\eqno(7.21)$$ with:
$$\delta R_{PT}(Q^2)=R_{reg,<}^{PT}(Q^2)- R_<^{PT}(Q^2)\eqno(7.22)$$ The ``perturbative'' power
corrections are then obtained by taking the low
$\mu^2$ expansion of  
$\dot{{\cal F}_R}({\mu^2\over Q^2})$ inside the corresponding integrals of finite support
$[0,\Lambda^2]$ in eq.(7.18) and (7.20) (note that, since $\mu^2<\Lambda^2<Q^2$, 
$\delta R_{PT}(Q^2)$ depends only on the ``low energy piece'' ${\cal F}_{(-)}$ of ${\cal F}_R$).
For instance, an analytic term $n\left({\mu^2\over Q^2}\right)^n$ (with $n>0$  integer) in the
low-$\mu^2$ expansion of
$\dot{{\cal F}_R}({\mu^2\over Q^2})$ contributes a power correction $ b_n^{PT} 
\left({\Lambda^2\over Q^2}\right)^n$ (with $b_n^{PT}$ given in eq.(4.13) ). The same result holds
if
$n\neq$ integer, in particular in the phenomenologically important case $n=1/2$ (``$1/Q$ power
corrections'' [35]). In the one-loop coupling case, the result eq.(4.25) for
$b_n^{PT}$ (which contains the necessary ambiguous imaginary part when
$n\neq$ integer) agrees with the Beneke-Braun formula eq.(6.29) (with ${\cal F}\rightarrow {\cal
F}_{(-)}$, see  a general derivation in Appendix B ) . This result is remarkable, since it does
not rely on a dispersion relation for ${\cal F}_{(-)}$: in particular, for $n$ integer, such an
analytic term may be a ``subtraction'' term, unrelated to the discontinuity of ${\cal F}_{(-)}$
(Appendix B). Similarly, a non-analytic term 
$n\left({\mu^2\over Q^2}\right)^n (c_n \ln {Q^2\over\mu^2} + {\bar d}_n)$ in $\dot{{\cal F}_R}$
($n$ integer) contributes a (log-enhanced) power correction: 
$$c_n \left({\Lambda^2\over Q^2}\right)^n \left(b_n^{PT} \ln
{Q^2\over\Lambda^2}+\bar{b}_n^{PT}\right) + {\bar d}_n b_n^{PT} \left({\Lambda^2\over
Q^2}\right)^n\eqno(7.23)$$ with $\bar{b}_n^{PT}=J_n - J_n^{PT}$, where:
$$J_n=\int_0^{\Lambda^2}n{d\mu^2\over\mu^2}\left({\mu^2\over \Lambda^2}\right)^n\
\ln{\Lambda^2\over\mu^2} \ \alpha_{eff}^{PT}(\mu^2)\eqno(7.24)$$ and:
\begin{eqnarray*}\ \ \ \ \ \ \ \ \ \ \ \ \ \ \ \ \ \ \ \ \ \ \ J_n^{PT}& = &\int_0^\infty dz\  
\tilde{\alpha}_{eff}(z)\ 
\left[\int_0^{\Lambda^2}n{d\mu^2\over
\mu^2}\
\left({\mu^2\over \Lambda^2}\right)^n\ \ln{\Lambda^2\over\mu^2}\ 
\exp\left(-z\beta_0\ln{\mu^2\over \Lambda^2 }\right)\right]\\ &=&{1\over n}\ \int_0^\infty dz\  
 \tilde{\alpha}_{eff}(z)\
\frac{1}{\left(1-\frac{z}{z_n}\right)^2}\ \ \ \ \ \ \ \ \ \
\ \ \ \ \ \ \ \ \ \
\ \ \ \ \ \ \ \ \ \
\ \ (7.25)\end{eqnarray*} is the Borel sum corresponding to $J_n$. The same techniques applied to
the representation eq.(2.17) of an Euclidean quantity $D(Q^2)$ reproduce the results of section
6.1, with explicit expressions similar to eq.(7.24),(7.25) for the coefficients of the subleading
log terms.  Note that $J_n^{PT}$, hence $\bar{b}_n^{PT}$, are ambiguous, due to the presence of
an IR renormalon (a simple pole) at $z=z_n$ in the Borel transform, the simple zero in
$\tilde{\alpha}_{eff}(z)$ only partially cancelling the double pole in the integrand of
eq.(7.25). This is another  example of the relation [4,6] between non-analytic terms in the
characteristic function and IR renormalons. This relation too can only be understood if
$\alpha_{eff}^{PT}(\mu^2)$ has a non-trivial IR fixed point: otherwise, if one assumes e.g.
$\alpha_{eff}^{PT}(\mu^2)$ is given by the one-loop coupling (i.e. $\tilde{\alpha}_{eff}(z)
\equiv 1$), one could associate IR renormalons even to analytic terms in the low $\mu^2$ expansion
of $\dot{{\cal F}_R}({\mu^2\over Q^2})$! 

\noindent On the other hand, the coefficients
$b_n^{PT}$ of the leading-log parts (and in particular of the analytic parts if there are no
accompanying log) are unambiguous  for $n$ integer (eq.(4.13)) if $\tilde{\alpha}(z)$ has no
renormalon, which suggests (in agreement with the analysis of section 6.1) they should be
associated to short-distances : this point is tricky, since  {\em all} power corrections formally
originate (see eq.(7.22)) from integration over {\em low}
$\mu^2$, and shows  it is misleading   to use the $\alpha_{eff}$ representations eq.(2.5b) or
(7.2) to separate long  from short distances  , at the difference (section 6) of the
$\alpha_s$ - representation eq.(2.1)  . 

\noindent The above interpretation is reinforced by the following observation:  the time-like
discontinuity of the {\em leading} log  contributions to
$\delta D_{PT}$ , which   arise (as discussed in section 6.1) from {\em short} distances, are
related to the corresponding {\em leading} log contributions to ${d(\delta R_{PT})\over d\ln
Q^2}$ by eq.(7.5). For instance, in the case of the Euclidean quantity
$D(Q^2)$ of eq.(6.8) with $n$ integer,   one gets for the associated (through eq.(7.12))
time-like observable:
$\delta R_{PT}(Q^2)= b_n^{PT}\left({\Lambda^2\over Q^2}\right)^n$ (which is {\em entirely} short
distance!), whereas the corresponding leading log contribution to
$\delta D_{PT}$ is ( see eq.(6.20)): $-n b_n^{PT} \ln {Q^2\over
\Lambda^2}  \left(-\ {\Lambda^2\over Q^2}\right)^n$, whose time-like discontinuity is indeed
related to $\delta R_{PT}$ by eq.(7.5). Similarly, for the quantity of eq.(6.24) , one gets
$\delta R_{PT}(Q^2)_{\mid leading\ log}= b_n^{PT}\left({\Lambda^2\over Q^2}\right)^n \ln
{Q^2\over \Lambda^2}$, whereas the corresponding leading log contribution to
$\delta D_{PT}$ is ( see eq.(6.25)): $- {1\over 2}n b_n^{PT} \ln^2 {Q^2\over
\Lambda^2}  \left(-\ {\Lambda^2\over Q^2}\right)^n$, whose time-like discontinuity is again
related to ${d(\delta R_{PT})\over d\ln Q^2}_{\mid leading\ log}$ by eq.(7.5).  The basic reason
for these relations is as follows.  The dispersion relation in eq.(7.16) implies the ``inverse''
relation [18,36]:
$$R_{reg}^{PT}(Q^2)={1\over 2\pi i} \oint_{|Q'^2|=Q^2}{dQ'^2\over Q'^2}\
D_{reg}^{PT}(Q'^2)\eqno(7.26)$$ But it is clear that, at least formally, the Borel sum $R_{PT}$
is also related to the Borel sum
$D_{PT}$ by eq.(7.26)  and
$R_{PT}$ may actually be obtained by substituting $D_{reg}^{PT}$ with $D_{PT}$ of eq.(6.5a) into
eq.(7.26), and permutting (as usual) the order of integrations! It follows the same statement is
true for
$\delta R_{PT}$ and
$\delta D_{PT}$, i.e. we have:
$$\delta R_{PT}(Q^2)={1\over 2\pi i} \oint_{|Q'^2|=Q^2}{dQ'^2\over Q'^2}\ \delta
D_{PT}(Q'^2)\eqno(7.27)$$ which accounts for the above mentionned relations . This  argument
remains formal as long as no precise definition of the ( ambiguous) Borel sums $D_{PT}$  and
$R_{PT}$ (hence of
$\delta D_{PT}$ and $\delta R_{PT}$) is given: the principal part prescription is adequate here,
since the previous argument is valid with it\footnote{This is a different prescription that the
one which leads to eq.(A.1)  and to a vanishing time-like discontinuity of  $\delta D_{PT}$; of
course, given any prescription for
$D_{PT}$ , one can always define the corresponding
$R_{PT}$ by requiring that it is related to $ D_{PT}$ by eq.(7.26), but in general this would
result in different prescriptions for the Borel sums $D_{PT}$ and $R_{PT}$!}\ .

\subsection{Non-perturbative power corrections} They are contained in (cf. eq.(2.10b)):
$$\delta R_{NP}(Q^2)=\int_0^\infty{d\mu^2\over\mu^2}\ \delta\alpha_{eff}^{NP} (\mu^2)\ \dot{{\cal
F}_R}({\mu^2\over Q^2})\eqno(7.28)$$ and are obtained [7] by taking the low $\mu^2$ expansion of
$\dot{{\cal F}_R}({\mu^2\over Q^2})$ inside the integral  (since 
$\delta\alpha_{eff}^{NP}(\mu^2)$ is assumed to be exponentially suppressed at large
$\mu^2$). For instance,  a non-analytic term $n\left({\mu^2\over Q^2}\right)^n (c_n \ln
{Q^2\over\mu^2} + {\bar d}_n)$ in $\dot{{\cal F}_R}$ ($n$ integer) contributes a log-enhanced
power correction:
$$c_n \left({\Lambda^2\over Q^2}\right)^n \left(b_n^{NP} \ln
{Q^2\over\Lambda^2}+\bar{b}_n^{NP}\right) + {\bar d}_n b_n^{NP} \left({\Lambda^2\over
Q^2}\right)^n\eqno(7.29)$$ where $b_n^{NP}$ is given in eq.(6.38), and:
$$\bar{b}_n^{NP}=\int_0^\infty n{d\mu^2\over\mu^2}\left({\mu^2\over \Lambda^2}\right)^n\
\ln{\Lambda^2\over\mu^2} \ \delta\alpha_{eff}^{NP}(\mu^2)\eqno(7.30)$$ Note that eq.(7.29) has
{\em exactly} the same structure as the corresponding contribution to
$\delta R_{PT}(Q^2)$ (eq.(7.23)) with the substitutions $b_n^{PT}\rightarrow b_n^{NP}$ and
$\bar{b}_n^{PT}\rightarrow \bar{b}_n^{NP}$! Again~, the leading log terms terms with a
coefficient $b_n^{NP}$ (an {\em analytic, integer} moment) should be associated to short
distances, while the sub-leading log terms, with a coefficient $\bar{b}_n^{NP}$ (a {\em non-
analytic} moment), are partly long distance. One sees once more it is  not possible, for $n$
integer, to disentangle unambiguously these two type of contributions (which get mixed once one
changes the scale
$\Lambda$ inside the log in eq.(7.29)), unless $c_n=0$ or $b_n^{NP}=0$. The exception is again
the case $n\neq$ integer, where there are no logs of UV origin.

\noindent The non-perturbative  power corrections in $R(Q^2)$ may also be derived from those in
the associated Euclidean quantity $D(Q^2)$  , since the dispersion relation eq.(7.4) and its
inverse eq.(7.26) hold between
$\delta D_{NP}(Q^2)$ and $\delta R_{NP}(Q^2)$ (note it is always possible to reconstruct a
$D(Q^2)$
 corresponding to a given $R(Q^2)$ using the relation eq.(7.12), so the  method below is general)
. In particular we have (eq.(7.5)):
$${d(\delta R_{NP})\over d\ln Q^2}=-\frac{1} {2\pi i} Disc\{\delta D_{NP}(-Q^2)\} \ \ \ \ \ \
(Q^2>0)\eqno(7.31)$$ Consider for instance $D(Q^2)$ in eq.(6.24) with $n$ integer, which is the
Euclidean quantity associated to:
$$R(Q^2)=\int_{0}^{Q^2}n{d\mu^2\over
\mu^2}\ \alpha_{eff}(\mu^2)\ \left(\mu^2\over Q^2\right)^n\ \ln{Q^2\over\mu^2}\eqno(7.32)$$ and
assume [7] $\delta\alpha_s^{NP}$ is restricted to low $k^2$, so that all $b_p^{NP}$'s vanish for
$p$ integer. Then one gets for $Q^2\gg\Lambda^2$:
$$\delta D_{NP}(Q^2)\simeq K_n^{NP}\ \left({\Lambda^2\over Q^2}\right)^n \ln {Q^2\over\Lambda^2}
+ const \eqno(7.33)$$ with $K_n^{NP}$ given in eq.(6.43), whereas (eq.(7.29)):
$$\delta R_{NP}(Q^2)\simeq \bar{b}_n^{NP} \left({\Lambda^2\over Q^2}\right)^n \eqno(7.34)$$
Eq.(7.31) then implies: 
$$K_n^{NP}=-(-1)^n n\ \bar{b}_n^{NP}\eqno(7.35)$$ i.e.:
$$\int_{0}^\infty{dk^2\over k^2}\ \left({k^2\over \Lambda^2}\right)^n\
\delta\alpha_s^{NP}(k^2) = (-1)^n n \int_0^\infty {d\mu^2\over\mu^2}\ \left({\mu^2\over
\Lambda^2}\right)^n\ \ln {\mu^2\over
\Lambda^2}\ 
\delta\alpha_{eff}^{NP} (\mu^2)\ \ \ \ \ \ (n = integer)\eqno(7.36)$$ Similarly, if  $n \neq$
integer and:
$$R(Q^2)=\int_{0}^{Q^2}n{d\mu^2\over
\mu^2}\ \alpha_{eff}(\mu^2)\ \left(\mu^2\over Q^2\right)^n\eqno(7.37)$$ the associated Euclidean
quantity is as in eq.(6.8), and one gets (eq.(6.42)) (again assuming the
$b_p^{NP}$'s vanish for integer $p$):
$$\delta D_{NP}(Q^2)\simeq K_n^{NP}\ \left({\Lambda^2\over Q^2}\right)^n \eqno(7.38)$$ while:
$$\delta R_{NP}(Q^2)\simeq b_n^{NP} \left({\Lambda^2\over Q^2}\right)^n \eqno(7.39)$$ In this
case eq.(7.31) implies:
$$K_n^{NP} = {\pi n\over\sin(\pi n)}\ b_n^{NP}\ \ \ \ \ \ (n \neq integer) \eqno(7.40)$$ (which
is the exact replic of eq.(A.16) of Appendix A!), i.e.:
$$\int_{0}^\infty{dk^2\over k^2}\ \left({k^2\over \Lambda^2}\right)^n\
\delta\alpha_s^{NP}(k^2) = {\pi n\over \sin\pi n}
\int_0^\infty{d\mu^2\over\mu^2}\ \left({\mu^2\over \Lambda^2}\right)^n\ \delta\alpha_{eff}^{NP}
(\mu^2)\ \ \ \ \ \ (n \neq integer)\eqno(7.41)$$ (eq.(7.41) is   valid for $0<n<1$  even if
$\delta\alpha_s^{NP}(k^2)= {\cal O}(\Lambda^2/k^2)$ at large $k^2$ since the integral on the left
hand side of eq.(7.41) is still UV convergent in this case).  

\noindent Alternatively, eq.(7.35) and (7.40) may be derived by comparing the two  expressions
for $\delta D_{NP}(Q^2)$ obtained from the equivalent representations eq.(2.10a) and (2.10b),
choosing
$D(Q^2$ as in eq.(6.8). Eq.(2.10a) yields eq.(7.38) . On the other hand, using eq.(6.31) and
(6.32) for $n$ integer and $0<n<1$ respectively, one gets from eq.(2.10b):
$$\delta D_{NP}(Q^2)\simeq -(-1)^n n\ \bar{b}_n^{NP} \ \left({\Lambda^2\over Q^2}\right)^n
\eqno(7.42)$$ if $n=$ integer, and:
$$\delta D_{NP}(Q^2)\simeq {\pi n\over\sin(\pi n)}\ b_n^{NP} \ \left({\Lambda^2\over Q^2}\right)^n
\eqno(7.43)$$ if $0<n<1$, which reproduce eq.(7.35) and (7.40) upon comparaison with eq.(7.38)~.
The relations eq.(7.36) and (7.41) may be useful, since the moments are treated in [7] as fit
parameters which constrain the shape of
$\delta\alpha_{eff}^{NP}(\mu^2)$, hence
$\delta\alpha_s^{NP}(k^2)$, and it may be easier to find a fit for the latter quantity then for
the former (which must be a complicated oscillating function to satisfy the constraint that its
integers moments $b_n^{NP}$ (eq.(6.38)) vanish).

\section {Applications}
\subsection{Hadronic width of the $\tau$ lepton}  It is usually expressed in term of the quantity
$R_{\tau}$, itself related to the total $e^+e^-$ annihilation cross-section into hadrons
$R_{e^+e^-}$ and to the Adler $D_{e^+e^-}$ function by:
\begin{eqnarray*}\ \ \ \ \ \ \ \ \ \ \ \  R_{\tau}(m_{\tau}^2)&=&2\int_0^{m_{\tau}^2}{ds\over
m_{\tau}^2}\left(1-{s\over m_{\tau}^2}\right)^2 \left(1+2{s\over m_{\tau}^2}\right)
R_{e^+e^-}(s)\\ & = &{1\over 2\pi i}\oint_{|s|=m_{\tau}^2}{ds\over s}\left(1-{s\over
m_{\tau}^2}\right)^3 \left(1+{s\over m_{\tau}^2}\right) D_{e^+e^-}(s)\ \ \ \
\ \ \ \ \ \
\ \ \ \ (8.1)\end{eqnarray*} In the dressed single gluon exchange approximation one has (with the
parton model normalization removed):
$$R_{e^+e^-}(Q^2) = 1 + \int_0^\infty{d\mu^2\over\mu^2}\ \alpha_{eff}(\mu^2)\ \dot{{\cal
F}}_{e^+e^-}({\mu^2\over Q^2})\eqno(8.2)$$ where ${\cal F}_{e^+e^-}$ has been computed in
[6,7,18]. Eq.(8.1) and (8.2) imply:
$$R_{\tau}(m_{\tau}^2)=1+\int_0^\infty{d\mu^2\over\mu^2}\ \alpha_{eff}(\mu^2)\ \dot{{\cal
F}}_{\tau}({\mu^2\over m_{\tau}^2})\eqno(8.3)$$ with [7,18]:
$$\dot{{\cal F}}_{\tau}({\mu^2\over m_{\tau}^2})=2\int_0^{m_{\tau}^2}{ds\over m_{\tau}^2}
\left(1-{s\over m_{\tau}^2}\right)^2 \left(1+2{s\over m_{\tau}^2}\right) \dot{{\cal
F}}_{e^+e^-}({\mu^2\over s})\eqno(8.4)$$ In particular, one finds [7] in the small $\mu^2$ limit:
$${\cal F}_{\tau}({\mu^2\over m_{\tau}^2})-{\cal F}_{\tau}(0)\simeq -d_1^{\tau}{\mu^2\over
m_{\tau}^2}+...\eqno(8.5)$$ with:
$$d_1^{\tau}={16\over 3\pi}(4-3\zeta(3))\eqno(8.6)$$ which implies a leading $1/m_{\tau}^2$ power
correction\footnote{While this paper was in writing, I learned about the article [37], where the
presence of $1/m_{\tau}^2$ terms arising from ``dispersive regularization'' of the coupling is
also pointed out. The implementation of this idea is however very different from the present one:
there it is applied directly to the ``effective charge'' defined by the
$D_{e^+e^-}$ function itself. I thank A. Kataev for bringing this reference to my attention.} \
of UV origin:
$$\delta R_{\tau}(m_{\tau}^2)\simeq b_1 d_1^{\tau} {\Lambda^2\over m_{\tau}^2}\eqno(8.7)$$
According to the discussion of section 7, this term originates directly from the integral on the
circle (eq.(8.1)) of a corresponding  leading UV ${\cal O}(1/Q^2)$ term  in $\delta
D_{e^+e^-}(Q^2)$ (such a term is present [6] in the large $\beta_0$ limit in  $\delta
D_{e^+e^-}^{PT}(Q^2)$).  For a numerical estimate, assume:
$$b_1 \simeq b_1^{PT}|one-loop = -{1\over \beta_0}$$ and take:
$\Lambda=\Lambda_V=2.3\Lambda_{\overline {MS}}$ (this choice of parameters corresponds to the
large $\beta_0$ estimate [6] for $\delta R_{\tau}^{PT}$). Then one gets (for 3 flavors), assuming
$\alpha_s^{\overline {MS}}(m_{\tau}^2)=0.32$:
$$\delta R_{\tau}(m_{\tau}^2)\simeq -0.063$$ which represents a sizable correction with respect
to the (principal-value) Borel sum estimate [6] (still in the large $\beta_0$ limit): $
R_{\tau}(m_{\tau}^2)-1\simeq 0.227$ , or to the experimental value [38] $
R_{\tau}(m_{\tau}^2)-1\simeq 0.20$.

\noindent Note also that a corresponding $1/Q^2$ power correction is absent from $R_{e^+e^-}(Q^2)$
(for which the leading power correction (of UV origin) is only ${\cal O}(1/Q^4)$ [7] ).

\subsection{Gluon condensate on the lattice}  Power corrections of UV origin may be relevant for
the lattice determination of the gluon condensate, since they are likely to affect the so-called
``perturbative tail''. To see this,  consider the following model [39] for the lattice plaquette
$W(\alpha)$:
$$Q^4\times W(\alpha)=\int_{0}^{Q^{2}}  dk^2 k^2
\alpha_s(k^2/Q^2,\alpha)\eqno(8.8)$$  where Q is the UV cutoff (of the order of the inverse
lattice size) and $\alpha$ the bare coupling constant. Eq.(8.8) is a peculiar case of eq.(6.8)
for $n=2$;  with the renormalized coupling 
$\alpha_s$ given by eq.(2.13), one expects the short distance expansion of $W(\alpha)$ to involve
${\cal O}(1/Q^2)$ contributions  which can screen the ``physical'', 
``genuine non-perturbative'' gluon condensate ${\cal O}(1/Q^4)$ contribution.~Indeed one gets:
$$W(\alpha)=W_{PT}(\alpha)+\delta W(\alpha)\eqno(8.9)$$  where $Q^4\times W_{PT}(\alpha)$ is the
Borel sum which defines the quartically divergent ``perturbative tail'' of eq.(8.8), while ( see
eq.(6.20) and Appendix A2):
$$Q^4\times \delta W(\alpha) = b_1 \Lambda^2 Q^2 - (b_2 \ln {Q^2\over \Lambda^2} + {\bar D}_2 \pm
i\pi b_2^{PT}) \Lambda^4 + {\cal O}(1/Q^2) \eqno(8.10)$$ is the subdominant ``power correction''
term, which involves still quadratic and logarithmic divergences. Eq.(8.10) suggests that, after
the ambiguous $\pm i\pi b_2^{PT}$ imaginary part has been absorbed into the Borel integral , the
first detected correction to the resulting (principal value regularized) Borel sum of the
perturbative tail may be the $b_1 \Lambda^2 Q^2$ quadratically divergent term (such a term  may 
even have been detected in a preliminary analysis [40] of the results of [39], where
$8$ orders of the perturbative expansion of $W_{PT}(\alpha)$ have been computed). Furthermore,
the UV finite ``gluon condensate'' is burried into the constant ${\bar D}_2$, which contains both
``perturbative'' and ``non-perturbative'' contributions from $\delta\alpha_s^{PT}$ and
$\delta\alpha_s^{NP}$ respectively (I have set ${\bar D}_2 \pm i\pi b_2^{PT}\equiv
D_2^{PT}+D_2^{NP}$). It appears however impossible to fix this constant independently of the
(arbitrary) choice of the scale $\Lambda$ inside the log divergence, and to separate
contributions of IR and UV origin in ${\bar D}_2$. The condition
$b_2=0$ (i.e. the absence of such a divergence) thus appears as a {\em minimal} requirement for an
unambiguous definition of the condensate  as a quantity of genuine IR origin. The situation here
appears even more severe then in the case
$b_2=0$, where  the definition of the condensate depends on the (partially arbitrary) [41]
definition of the ``regularized sum'' (through principal value prescription or else) [21,22] of
perturbation theory, and its extraction from lattice data already faces  serious difficulties
[42].

\section{Conclusion} In this paper, I have given  arguments to support the existence, within the
dispersive approach [7], of power corrections, some of them {\em of short distance origin} 
(hence unrelated to - thus {\em not inconsistent with} - the OPE),  which appear naturally when
the Landau singularity in the running coupling is removed. For an euclidean observable $D(Q^2)$,
the situation can be summarized as follows: introducing an IR cut-off at $k^2=\Lambda_I^2$ as in
section 6, one can split eq.(2.1) into a ``long distance'' and a ``short distance'' part:
$$D(Q^2) =\int_{0}^{\Lambda_I^2}{dk^2\over k^2}\ \alpha_s(k^2)\ \varphi\left(k^2\over Q^2\right)
+ \int_{\Lambda_I^2}^\infty{dk^2\over k^2}\ \alpha_s^{PT}(k^2)\ \varphi\left(k^2\over Q^2\right)
+ \int_{\Lambda_I^2}^\infty{dk^2\over k^2}\ \delta\alpha_s(k^2)\ \varphi\left(k^2\over
Q^2\right)\eqno(9.1)$$ which represents an example of OPE ``a la SVZ'' [30-32]. In the short
distance part, I have further split the IR regular coupling $\alpha_s$ into a``perturbative'' and
a ``power correction'' piece (eq.(2.13)).  The long distance part yields, for large $Q^2$, power
corrections, which one can parametrize [8] with the IR regular coupling, and are  consistent with
the OPE. The integral over the perturbative coupling in the short distance part represents a form
of ``regularized perturbation theory '' [21,22] (choosing the IR cut-off
$\Lambda_I$ above the  Landau singularity
$\Lambda$ of $\alpha_s^{PT}$). The last integral in eq.(9.1) yield at large $Q^2$ the new power
corrections of short distance origin discussed in this paper, unrelated to the OPE. Equivalently,
eq.(9.1) can be rewritten as:
$$D(Q^2) = \int_0^\infty{dk^2\over k^2}\ \alpha_s^{PT}(k^2)\ \varphi\left(k^2\over Q^2\right) +
\int_{0}^{\Lambda_I^2}{dk^2\over k^2}\ \delta\alpha_s(k^2)\ \varphi\left(k^2\over Q^2\right) +
\int_{\Lambda_I^2}^\infty{dk^2\over k^2}\ \delta\alpha_s(k^2)\ \varphi\left(k^2\over
Q^2\right)\eqno(9.2)$$ which shows that the (ambiguous) Borel sum of perturbation theory (the
first integral on the right hand side of eq.(9.2)) receive two types of power corrections at
large $Q^2$: the long distances ones (the second integral), which correspond to the standard OPE
``condensates'' contribution [20] (and contain both an ambiguous, ``perturbative'' component
coming from the 
$\delta\alpha_s^{PT}$ piece of $\delta\alpha_s$ and a ``genuine non-perturbative'' component from
the
$\delta\alpha_s^{NP}$ piece); and those arising from short distances (the last integral) . If the
short distance power corrections are neglected [8] (i.e. if one assumes that
$\delta\alpha_s(k^2)$ is essentially a low $k^2$ modification and decreases sufficiently fast  at
large
$k^2$), one recovers the standard view (see e.g. [6], [22]) that the first correction to the
Borel sum is given by the OPE. 

\noindent Since the
$\delta\alpha_s^{PT}$ piece  ( which eliminates the Landau singularity ) is however a priori {\em
not} restricted to low $k^2$, it induces ``perturbative power corrections''
$\delta D_{PT}$  which  arise  both from long distances (where they remove the IR renormalons
ambiguities of the Borel sum) and from short distances. They stand on the same level as radiative
corrections, and  are best looked upon as part of the ``correct'' resummation prescription of
perturbation theory, yielding a ``regularized perturbation theory '' [21,22]  which differs from
Borel summation (even barring IR renormalons problems). There is no contradiction with the OPE,
which does not require that {\em all} power contributions be of long distance origin. 

\noindent The occurence of ``non-OPE'', ``short-distance'' power contributions is even more
conspicuous from the ``$\alpha_{eff}$ representation''  eq.(2.17). Performing an analogue split at
$\mu^2=\Lambda^2$ as in section 7.1, one gets:
$$D(Q^2)\simeq \int_0^{\Lambda^2}{d\mu^2\over\mu^2}\ \alpha_{eff}(\mu^2)\ \dot{{\cal
F}}({\mu^2\over Q^2})+\int_{\Lambda^2}^\infty{d\mu^2\over\mu^2}\ \alpha_{eff}^{PT}(\mu^2)\
\dot{{\cal F}}({\mu^2\over Q^2})\eqno(9.3)$$ where $\alpha_{eff}\equiv
\alpha_{eff}^{PT}+\delta\alpha_{eff}^{NP}$ (both contributions are assumed to be IR regular), and
the
$\delta\alpha_{eff}^{NP}$ piece has been neglected in the high $\mu^2$ integral, since I assume
it is exponentially supressed there. Again, the low $\mu^2$ integral generates  power
contributions at large $Q^2$, which one can parametrize [8] with low  $\mu^2$ moments of the {\em
total} IR regular effective coupling $\alpha_{eff}$ (after expanding $\dot{{\cal F}}$ ). But it
is clear that  any term in the  low  $\mu^2$ expansion of $\dot{{\cal F}}$ (either analytic or
non-analytic) can a priori contribute a power correction, whereas only the non-analytic terms are
related (section 6) to OPE and long distances [4,6,7]. It seems artificial to eliminate the
analytic contributions, which are associated to short distances [4,6], and require the first
few\footnote{This condition cannot be imposed [33] on {\em all} of them,  since they are defined
on a {\em finite} interval.} \  analytic low  $\mu^2$ moments of
$\alpha_{eff}$ to vanish; furthermore, they certainly cannot vanish if, as is likely,
$\alpha_{eff}$ remains positive at low scales ( this requirement looks more plausible, as argued
below, if one postulate it [7] for the
$\delta\alpha_{eff}^{NP}$ piece only)!

\noindent I should stress that the short distance ``perturbative power corrections'' here
discussed should not be confused with the effective $1/Q^2$ power correction which represents the
estimate [32,43,44] of the effect of UV renormalons on the remainder of the Borel sum when the
perturbative expansion is trunkated at its minimum term: even if the Borel summation is performed
exactly [36] (using, say, a principal part prescription) the short distance $1/Q^2$ power
correction in
$\delta D_{PT}$ still remains  ( moreover they are also present in observables with an UV cut-off
at $Q^2$, such as
$D(Q^2)$ in eq.(6.8), or the lattice plaquette (section 8.2), which do not have any UV
renormalon!).

\noindent The occurence of power corrections to the Borel sum arising from ``IR regularized''
couplings has been noted before in
 [6] (the resulting resummation for Minkowskian observables has also been considered in [18]) .
The point of view put forward here however departs from the one in [6] , which disfavor the use
of IR regular couplings such as $\alpha_{s,reg}^{PT}$ on the ground of the {\em assumption}
[6,22] that the leading power correction to the (Borel-summed) perturbation theory should be
given by the OPE: as argued in this paper, the framework of [7], although not in contradiction
with the OPE, does suggest the opposite assumption as a natural alternative, since the dispersive
regularization [10-13] of the coupling generates ``for free'' (eq.(2.17)) the ``minimal'' power
corrections necessary to remove the Landau singularity. Furthermore, there is then no a priori
reason that the genuine ``non-perturbative '' modification
$\delta\alpha_s^{NP}(k^2)$  be   itself restricted to low $k^2$, i.e. that the $b_n^{NP}$'s
(eq.(6.38)) vanish. The split eq.(2.14) of
$\delta\alpha_s$ into a ``perturbative'' and a ``non-perturbative '' component then becomes to
some extent a matter of convention, and such is the split eq.(6.40) of the total $b_n$. 

\noindent If one still insists on implementing the notion of ``low energy modification '' , two
natural options appear. The first one assumes it should concern only the
$\delta\alpha_s^{NP}$ part of the coupling; this choice leads to the picture of [7], where
$b_n^{NP}=0$ and the``genuine'' non-perturbative power corrections are always consistent with the
OPE, but where  ``perturbative'' power corrections from $\delta\alpha_s^{PT}$ foreign to the OPE
could remain. The proper definition of $\delta\alpha_s^{PT}$  becomes a physical question, rather
then a matter of convention, since $\delta\alpha_s\simeq\delta\alpha_s^{PT}$ at large
$k^2$ in this case. The alternative (more artificial in my opinion) is to have the {\em total}
$\delta\alpha_s$ restricted to low $k^2$, which would make the present framework consistent with
the standard view as explained above. This option requires that $b_n^{NP}=-b_n^{PT}$, so either
the condition [7]
$b_n^{NP}=0$ has to be relaxed, or a proper redefinition of ``$\delta\alpha_s^{PT}$'' and
``$\delta\alpha_s^{NP}$''  has to be found (by reshuffling part of $\delta\alpha_s^{NP}$ into the
new ``$\delta\alpha_s^{PT}$''), such that ``$b_n^{NP}$''$=$ ``$b_n^{PT}$'' $=0$. Whether this can
be achieved in a unique way at all is not clear. The point of view adopted here is that such
questions should be decided by the data, rather then imposed a priori, and the $b_n$'s considered
as free parameters .

\noindent An important qualitative difference between the  ``perturbative'' and the 
``non-perturbative'' power corrections, which could help defining a ``correct'' splitting of
$\delta\alpha_s$, concerns the notion of ``mismatch''[20] between radiative and power
corrections. In many processes, it has been (apparently successfully) postulated [20] that the
formally leading ${\cal O}(\alpha_s)$ radiative corrections are actually numerically negligible 
(for $Q^2$ low enough, but still high enough to have convergence of the short distance expansion
in inverse powers of $Q^2$) compared to the genuine ``non-perturbative'' power corrections. This
``mismatch'' is likely to be absent for the ``perturbative'' power corrections here considered,
which are presumably of the same size as the radiative corrections. For instance, in the case of
the regularized one-loop coupling eq.(3.1), one finds that for
$k^2\gsim 2 \Lambda^2$ the ``radiative term'' $\alpha_s^{PT}(k^2) = {1\over\beta_0
\ln{k^2\over\Lambda^2}}$ is of comparable size to the first ``perturbative'' power correction
${1\over\beta_0} {\Lambda^2\over k^2}$! Note also that ``mismatch'' between radiative and
``non-perturbative''  power corrections then requires also  $\delta R_{PT}(Q^2)\ll\delta
R_{NP}(Q^2)$ for
$Q^2\gsim\Lambda^2$ (even if $\delta\alpha_s^{PT}(k^2)\gg\delta\alpha_s^{NP}(k^2)$ at large
$k^2$), which gives a rough justification to the simultaneous neglect of radiative and
``perturbative'' power corrections implicitly performed in standard QCD sum rules analysis. It
would be interesting to investigate the constraints on $\delta\alpha_s^{NP}$ necessary to
implement the ``mismatch''.

\noindent How unique is the present proposal ? The choice of Borel summation to define the
``unregularized'' sum of perturbation theory is as a rather natural one [45]. However , there is
still considerable freedom in the definition of $\alpha_{s,reg}^{PT}$ and
$\delta\alpha_s^{PT}$. For instance, as an alternative to eq.(3.10), one might consider
shopping-off the time-like discontinuity of the perturbative coupling  at
$\mu^2=c^2\Lambda^2$  instead of $\mu^2=0$:
$$\alpha_{s,reg}^{PT}(k^2)  = -\int_{c^2\Lambda^2}^\infty{d\mu^2\over\mu^2+k^2}\
\rho_{PT}(\mu^2)\eqno(9.4)$$ or even define $\alpha_{s,reg}^{PT}(k^2)\equiv \alpha_>^{PT}(k^2)$
(eq.(A.21)). The choice
$\delta\alpha_{eff}^{PT}\equiv 0$ in eq.(2.16) appears still a very natural  one, since one only
disturbs in a minimal way the information contained in perturbation theory. Of course, as long as
it is only a question of definition, it matters little what one calls ``perturbative'' and
``non-perturbative''; but the distinction becomes a meaningful one once one starts postulating
specific physical properties (such as ``low-energy restriction'' or ``mismatch'') for  any of
these pieces. 

\noindent Another  possible limitation of the present proposal is the condition of exponential
suppression of $\delta\alpha_{eff}(\mu^2)$: this excludes such  familiar model as the
``Richardson-like'' type of IR finite coupling:
$$\alpha_s(k^2) = {1\over\beta_0\ln\left(c^2+k^2/\Lambda^2\right)}\eqno(9.5)$$ ($c^2\geq 1$)
where the large $k^2$ expansion of $\delta\alpha_s$ contains logs. Moreover, the possibility that
$\alpha_s^{PT}$ might itself be afflicted by renormalons ambiguities is worrysome, since it makes
the whole scheme more cumbersome, and with a less definite starting point. Nevertheless, given
the generic possibility of power corrections unrelated to the OPE arising from the idea of a
universal IR regular QCD coupling, it would be worthwhile to  develop practical ways to  get
phenomenological evidence for presence or absence of such terms.
\\

\noindent{\bf Acknowledgements}\\
 
\noindent I thank G. Altarelli, M. Beneke,  Yu.L. Dokshitzer, G. Marchesini and N.G. Uraltsev for
stimulating discussions, and V.I.Zakharov for an early explanation of the notion of mismatch.\\
\newpage

\noindent{\bf Appendices}\\

\noindent{\bf A \ \ \ \ \  Some  results on the dispersively regularized coupling}\\

\noindent{\em Assuming} $\alpha_s^{PT}(k^2)$ has {\em no complex singularities}, it was shown in
section 7  that in the time-like region ($Q^2>0$)
$Disc\{D_{reg}^{PT}(-Q^2)\}=Disc\{D_{PT}(-Q^2)\}$, which reflects the basic assumption of the
dispersive regularization method that
$Disc\{\alpha_{s,reg}^{PT}(-\mu^2)\}\equiv Disc\{\alpha_s^{PT}(-\mu^2)\}$. This observation,
which can equivalently be expressed as:
$$Disc\{\delta D_{PT}(-Q^2)\}\equiv 0\ \ \ \ \ \ (Q^2>0)\eqno(A.1)$$ has two interesting
consequences, when applied to the observable of eq.(6.8): i) it leads to a relation between the
$b_n^{PT}$'s and the IR renormalons residues in $\tilde{D}(z)$, and ii) it implies the
universality of the one-loop value of the IR fixed point:
$$\alpha_{s,reg}^{PT}(k^2=0)=\alpha_{eff\mid IR}^{PT}={1\over \beta_0}\eqno(A.2)$$ which is valid
for a general coupling, {\em provided $\tilde{\alpha}_s(z)$ has no renormalons, and
$\alpha_s^{PT}(k^2)$ has no complex singularities  and vanishes for $k^2\rightarrow 0$}.\\

\noindent{\bf A1 \ \ \ \ \  Universality of $\alpha_{eff\mid IR}^{PT}$ }\\

Let us first consider point ii). If $0<n<1$, eq.(A.1) determines the phase of $K_n^{PT}$
(eq.(6.27)) as:
$$ K_n^{PT}= (-1)^n  K_n\eqno(A.3)$$ with $K_n$ {\em real}, in order to give in eq.(6.26) a
contribution:
$$\delta D_{PT}(Q^2)\supset  K_n \left(-\ {\Lambda^2\over Q^2}\right)^n\eqno(A.4)$$ {\em
unambiguous} in the time-like region ($Q^2<0$).  Since the imaginary part of $ K_n^{PT}$ must
cancell the one in $D_{PT}(Q^2)$ generated by the IR renormalon, one deduces that in the {\em
space-like} region ($Q^2>0$):
$${1\over\pi} Im D_{PT}(Q^2) = K_n{\sin(\pi n)\over\pi} \left({\Lambda^2\over Q^2}\right)^n \ \ \
\ \ (0<n<1)\eqno(A.5)$$ On the other hand,  ${1\over\pi} Im D_{PT}(Q^2)$ is related (for any $n$)
to the IR renormalon residue ${\bar K}_n$ of the {\em ordinary} Borel transform $D(z)$ (in e.g.
the scheme where the {\em inverse} $\beta$ function has only two terms) by [22,23]:
$${1\over\pi} Im D_{PT}(Q^2) = \left[{\bar K}_n\ {1\over \Gamma (1+\delta)}\
z_n^{1+\delta}\right]\left({{\bar\Lambda}^2\over Q^2}\right)^n\eqno(A.6)$$ where, for $z
\rightarrow z_n $ (assuming $\tilde{\alpha}_s(z)$ has no renormalon):  
$$D(z) \simeq  \frac{{\bar K}_n}{(1 - \frac{z}{z_n})^{1+\delta}}\eqno(A.7)$$ with [30] $\delta
={\beta_1\over \beta_0} z_n$,  and ${\bar\Lambda}/\Lambda$ is an $n$-independant constant.
Eq.(A.7) follows e.g. from the  relation [27] between ordinary and modified Borel transforms
singularities and eq.(6.5b),  which yields the {\em RS invariant} Borel transform:
$$\tilde{D}(z) = \tilde{\alpha}_s(z)\ \frac{1}{1-\frac{z}{z_n}}\eqno(A.8)$$ hence, for $z
\rightarrow z_n $ (if
$\tilde{\alpha}_s(z)$ has no renormalon):
$$\tilde{D}(z) \simeq  \frac{{\tilde K}_n}{1 - \frac{z}{z_n}}\eqno(A.9)$$ with:
$${\tilde K}_n=\tilde{\alpha}_s(z_n)\eqno(A.10)$$ Comparing eq.(A.5) and (A.6), and letting
$n\rightarrow 0$, one gets:
$$K_0={{\bar K}_0\over \beta_0}={1\over \beta_0}\eqno(A.11)$$ where the last step follows also
from  [27]  which  implies, for $n\rightarrow 0$ (hence
$z_n\rightarrow 0$), that ${\bar K}_0=\tilde{K}_0$, hence:
$${\bar K}_0=\tilde{\alpha}_s(z=0)=1\eqno(A.12)$$
  
\noindent On the other hand, eq.(6.27) and (A.3) imply:
$$K_0=K_0^{PT}= \left[\int_0^\infty n{dk^2\over k^2}\left({k^2\over\Lambda^2}\right)^n
\delta\alpha_s^{PT}(k^2)\right]_{n\rightarrow 0}  = \alpha_{eff\mid IR}^{PT}\eqno(A.13)$$ where
the last equality is a consequence ot the assumption that $\alpha_s^{PT}(k^2=0)=0$, since then
$\delta\alpha_s^{PT}(k^2=0)=\alpha_{eff\mid IR}^{PT}$ (eq.(3.15)), and of the observation that
the integral in eq.(A.13) is dominated for $n\rightarrow 0$ by the $k^2\rightarrow 0$ region 
(where it is IR divergent). Eq.(A.11) and (A.13)  prove eq.(A.2). This argument explains and
extends to all orders the stability of 
$\alpha_{eff\mid IR}^{PT}$ with respect to higher order corrections pointed out in [12] (where a
general result has also been announced). Note it is crucial that
$\tilde{\alpha}_s(z)$ has no renormalons, and 
$\alpha_s^{PT}(k^2)$   no complex singularities. Otherwise, one could start from an arbitrary
$\alpha_{eff}^{PT}(\mu^2)$, with an arbitrary $\alpha_{eff\mid IR}^{PT}$, and reconstruct
$\alpha_{s,reg}^{PT}(k^2)$ via the dispersion relation eq.(3.10b). But the resulting
$\tilde{\alpha}_s(z)$ would in general have renormalons (from eq.(4.6)), and even if this is
avoided by having $\tilde{\alpha}_{eff}(z)$ vanish at $z=z_n$,  it is not guaranteed the
resulting $\alpha_s^{PT}(k^2)$ from eq.(4.1) has no complex singularities, or vanishes at
$k^2=0$! These conditions are in particular satisfied if $\alpha_s^{PT}(k^2)$ is the sum of an
arbitrary {\em finite} number of two-loop 't Hooft couplings (eq.(3.5)). \\

\noindent{\bf A2 \ \ \ \ \   Relation between $b_n^{PT}$ and the IR renormalons residues}\\
 
\noindent To prove   point i), assume first  $n$ is an integer. Then eq.(A.1) requires similarly
that the ambiguous imaginary part of the  constant $ D_n^{PT}$ in eq.(6.20) is given by $\pm i\pi
b_n^{PT} 
\left(-\ {\Lambda^2\over Q^2}\right)^n$, so that it can be merged with the log to give a
contribution:
$$\delta D_{PT}(Q^2)\supset -nb_n^{PT} 
\ln \left(-\ {Q^2\over\Lambda^2}\right)  \left(-\ {\Lambda^2\over Q^2}\right)^n\eqno(A.14)$$ 
{\em unambiguous} in the time-like region ($Q^2<0$). Since the imaginary part of
$ D_n^{PT}$ must cancell the one in $D_{PT}(Q^2)$ generated by the IR renormalon, one deduces
that in the {\em space-like} region ($Q^2>0$):
$${1\over\pi} Im D_{PT}(Q^2) = nb_n^{PT} \left(-\ {\Lambda^2\over Q^2}\right)^n\eqno(A.15)$$
which relates $b_n^{PT}$ to the $z=z_n$ IR renormalon residue of the standard   integral eq.(6.8)
through eq.(A.6) (eq.(A.15) also suggests  the  IR renormalon residue should  vanish if
$b_n^{PT}= 0$, so that there should be no discontinuity neither in the space-like nor in the
time-like region). Furthermore, eq.(A.15) is also valid for  $n\neq$  integer (defining a {\em
complex}
$b_n^{PT}$ either through eq.(3.14) or eq.(4.13)). This statement follows immediately from
eq.(A.3) and (A.5) and the relation:
$$ K_n^{PT} = {\pi n\over\sin(\pi n)}\ b_n^{PT}\ \ \ \ \ \ (0<n<1) \eqno(A.16)$$ To prove the
latter, one can either use eq.(3.12) into eq.(6.27) and compare to eq.(3.14) with the help of the
identity eq.(6.33), or start from  the relation  :
$$\delta\alpha_s^{PT}(k^2)=\alpha_{reg,<}^{PT}(k^2) - \alpha_<^{PT}(k^2)\eqno(A.17)$$ with:
$$\alpha_{reg,<}^{PT}(k^2)\equiv k^2\int_0^{\Lambda^2}{d\mu^2\over(\mu^2+k^2)^2}\
\alpha_{eff}^{PT}(\mu^2)\eqno(A.18)$$ and:
$$\alpha_<^{PT}(k^2)\equiv \int_0^\infty dz\ \tilde{\alpha}_{eff}(z)\left[k^2\int_0^ {\Lambda^2}
{d\mu^2\over(\mu^2+k^2)^2}\ \exp\left(-z\beta_0\ln{\mu^2\over
\Lambda^2 }\right)\right]\ \ \ \ (k^2>\Lambda^2)\eqno(A.19)$$ To derive eq.(A.17), one splits the
dispersive integral eq.(3.10b) at $\mu^2=\Lambda^2$, and write:
$$\alpha_{s,reg}^{PT}(k^2)=\alpha_{reg,<}^{PT}(k^2)+\alpha_>^{PT}(k^2)\eqno(A.20)$$ where:
$$\alpha_>^{PT}(k^2)\equiv k^2\int_{\Lambda^2}^\infty{d\mu^2\over(\mu^2+k^2)^2}\
\alpha_{eff}^{PT}(\mu^2)\eqno(A.21)$$  and uses the   Borel representation eq.(4.4) of
$\alpha_{eff}^{PT}(\mu^2)$ to get:
$$\alpha_>^{PT}(k^2)=\alpha_s^{PT}(k^2)-\alpha_<^{PT}(k^2)\eqno(A.22) $$ Assume now $0<n<1$. Upon
insertion of eq.(A.17), eq.(6.27) becomes:
$$  K_n^{PT} = \int_0^\infty n{dk^2\over k^2}\left({k^2\over\Lambda^2}\right)^n
\alpha_{reg,<}^{PT}(k^2)-\int_0^\infty n{dk^2\over k^2}\left({k^2\over\Lambda^2}\right)^n
\alpha_<^{PT}(k^2)\eqno(A.23)$$ But, permutting the $k^2$ and $\mu^2$ integrations, one gets:
\begin{eqnarray*}\int_0^\infty n{dk^2\over k^2}\left({k^2\over\Lambda^2}\right)^n
\alpha_{reg,<}^{PT}(k^2)&=&\int_0^{\Lambda^2}n{d\mu^2\over\mu^2}\left({\mu^2\over
\Lambda^2}\right)^n\alpha_{eff}^{PT}(\mu^2)\left[\mu^2\int_0^
\infty {dk^2\over(\mu^2+k^2)^2}\ \left({k^2\over\mu^2}\right)^n\right]\\ &=&{\pi n\over\sin(\pi
n)}\ I_n\ \ \ \ \ \ \ \ \ \ \ \ \ \ \ \ \ \ \ \ \ \ \
\ \ \ \ \ \ \ \ \ \ \ \ \ \ \ \ \ \ \ \ \ \ \ \ (A.24)\end{eqnarray*} whereas:
\begin{eqnarray*}\int_0^\infty n{dk^2\over k^2}\left({k^2\over\Lambda^2}\right)^n
\alpha_<^{PT}(k^2)&=&\int_0^\infty dz\ \tilde{\alpha}_{eff}(z)\ \int_0^{\Lambda^2}n{d\mu^2\over
\mu^2}\left({\mu^2\over\Lambda^2}\right)^n\ \exp\left(-z\beta_0\ln{\mu^2\over
\Lambda^2 }\right)\\ &\times &\left[\mu^2\int_0^\infty {dk^2\over(\mu^2+k^2)^2}\
\left({k^2\over\mu^2}\right)^n\right]\\ &=&{\pi n\over\sin(\pi n)}\ I_n^{PT}\ \ \ \ \ \ \ \ \ \ \
\ \ \ \ \ \ \ \ \ \
\ \ \ \ \ \ \ \ \ \ \ \ \ \ \ \ \ \ \ \ \ \ \ \ \ \ (A.25)\end{eqnarray*} which, together with
eq.(4.13), proves eq.(A.16). In deriving eq.(A.25), I have freely used the Borel representation
(eq.(A.19)) of $\alpha_<^{PT}(k^2)$ inside the integral on the left hand side of eq.(A.25) down
to $k^2=0$ (although it is valid, similarly to eq.(4.1), only for
$k^2>\Lambda^2$), and permutted the order of $k^2$ and $z$ integrations.  This procedure is 
similar to the one which gives the ``correct'' result eq.(6.5) for the integral eq.(6.3) over
$\alpha_s^{PT}(k^2)$, and its justification [23] is essentially the same: namely
$\alpha_<^{PT}(k^2)$ also contains a Landau singularity (which cancells (eq.(A.22)) the one in
$\alpha_s^{PT}(k^2)$, since $\alpha_>^{PT}(k^2)$ is IR regular).

\noindent Note that, comparing eq.(A.15) and (A.6), and letting $n\rightarrow 0$, one gets:
$$b_0^{PT}={{\bar K}_0\over \beta_0}={1\over \beta_0}\eqno(A.26)$$ But assuming 
$\alpha_s^{PT}(k^2=0)$ vanishes implies (eq.(3.15)) $b_0^{PT}=\alpha_{eff\mid IR}^{PT}$  ( this
relation can also be derived from eq.(4.13) ) , which gives an alternative  proof of eq.(A.2).\\

\noindent{\bf B \ \ \ \ \   Expressions for power corrections in term of $\delta\alpha_s$}\\

\noindent Although a Minkowskian quantity $R(Q^2)$  cannot be parametrized directly in term of
$\alpha_s$ through a real integral representation similar to eq.(2.1),  it may be interesting to
point out that simple expressions do exist for the power corrections
$\delta R_{PT}$ and $\delta R_{NP}$ themselves in term of $\delta\alpha_s^{PT}$ and
$\delta\alpha_s^{NP}$ respectively. In the former case, they lead to
 an alternative proof of the Beneke-Braun formula [6] .\\  

\noindent{\bf B1 \ \ \ \ \   Perturbative power corrections}\\

\noindent The result follows from a comparaison of the similar expressions eq.(7.22) and (A.17)
for $\delta R_{PT}$ and $\delta\alpha_s^{PT}$. As noted in section 7, for $Q^2>\Lambda^2$ 
$\delta R_{PT}(Q^2)$ depends only on ${\cal F}_{(-)}({\mu^2\over Q^2})$, the ``low energy''
characteristic function (eq.(7.1)). I shall assume the latter satisfies a (subtracted) dispersion
relation. In general subtractions terms are  present, since there is no constraint on the high
$\mu^2$ behavior of ${\cal F}_{(-)}$. 

\noindent i) Consider first the  case of one subtraction at
$\mu^2=0$: then ${\cal F}_{(-)}$ is given by (see eq.(6.30)):
$${\cal F}_{(-)}({\mu^2\over Q^2})-{\cal F}_{(-)}(0) = \int_{0}^\infty{dk^2\over k^2}\
{-{\mu^2\over k^2}\over1+{\mu^2\over k^2}}\ \varphi_{(-)}\left(k^2\over Q^2\right)\eqno(B1)$$
Taking the $\mu^2$ derivative and inserting the result into the expressions eq.(7.18) and (7.20)
for $R_{reg,<}^{PT}$ and $R_<^{PT}$ one gets immediately, after permutting the integrals and using
eq.(A.18):
$$R_{reg,<}^{PT}(Q^2)=\int_{0}^\infty{dk^2\over k^2}\ \alpha_{reg,<}^{PT}(k^2)\
\varphi_{(-)}\left(k^2\over Q^2\right)\eqno(B.2)$$ whereas, using eq.(A.19) down to $k^2=0$ (this
is again justified because
$\alpha_<^{PT}(k^2)$ contains a Landau singularity) one finds: 
$$R_<^{PT}(Q^2)=\int_{0}^\infty{dk^2\over k^2}\ \alpha_<^{PT}(k^2)\ \varphi_{(-)}\left(k^2\over
Q^2\right)\eqno(B.3)$$ Subtracting eq.(B.3) from eq.(B.2) one ends up with:
$$\delta R_{PT}(Q^2)=\int_{0}^\infty{dk^2\over k^2}\ \delta\alpha_s^{PT}(k^2)\
\varphi_{(-)}\left(k^2\over Q^2\right)\eqno(B.4)$$

\noindent ii) Assume next two subtractions. Then:
$${\cal F}_{(-)}({\mu^2\over Q^2})-{\cal F}_{(-)}(0) = a_0{\mu^2\over Q^2} + \int_{0}^\infty 
{dk^2\over k^2}\ \left({-{\mu^2\over k^2}\over1+{\mu^2\over k^2}}+{\mu^2\over k^2}\right)\
\varphi_{(-)}\left(k^2\over Q^2\right)\eqno(B.5)$$ where:
$$a_0\equiv {\cal F}_{(-)}'(0)$$ (Eq.(B.1)  and (B.5) are analogues of eq.(2.6); the
discontinuity $\varphi_{(-)}\left(k^2\over Q^2\right)$ can however no more be interpreted as a
Feynman diagram kernel, and neither should the subtraction terms necessarily  be identified to
the real contribution, and the dispersive integral to the virtual one). Following the same steps
as in i), one gets:
$$R_{reg,<}^{PT}(Q^2)=-a_0 I_1 {\Lambda^2\over Q^2} + \int_{0}^\infty{dk^2\over k^2}\
\left[\alpha_{reg,<}^{PT}(k^2)-I_1 {\Lambda^2\over k^2}\right]\ \varphi_{(-)}\left(k^2\over
Q^2\right)\eqno(B.6)$$ where $I_1$ is defined in eq.(4.14). Similarly, one finds:
$$R_<^{PT}(Q^2)=-a_0 I_1^{PT} {\Lambda^2\over Q^2} + \int_{0}^\infty{dk^2\over k^2}\
\left[\alpha_<^{PT}(k^2)-I_1^{PT} {\Lambda^2\over k^2}\right]\ \varphi_{(-)}\left(k^2\over
Q^2\right)\eqno(B.7)$$ where $I_1^{PT}$ is defined in eq.(4.15). One deduces:
$$\delta R_{PT}(Q^2)=-a_0 b_1^{PT} {\Lambda^2\over Q^2} + \int_{0}^\infty{dk^2\over k^2}\
\left[\delta\alpha_s^{PT}(k^2)-b_1^{PT} {\Lambda^2\over k^2}\right]\ \varphi_{(-)}\left(k^2\over
Q^2\right)\eqno(B.8)$$ with $b_1^{PT}$ as in eq.(4.13). The subtraction term in the integrand 
insures convergence of the integral at infinity since $\delta\alpha_s^{PT}(k^2) \simeq b_1^{PT}
{\Lambda^2\over k^2}+{\cal O}({\Lambda^4\over k^4})$ (eq.(3.13)). Generalization to an arbitrary
number of subtractions is clear from eq.(B.4) and (B.8), which are analogues of eq.(6.4) for
Minkowskian observables. Note that non-analytic terms in the large $Q^2$ expansion of
$\delta R_{PT}(Q^2)$ and renormalons come only from the dispersive integral in eq.(B.5), and
 are in one-to-one correspondance with terms in the low energy expansion of
$\varphi_{(-)}\left(k^2\over Q^2\right)$.

\noindent iii) Extension to the general case where subtractions away from $\mu^2=0$ are required
proceeds along similar lines.  Assuming e.g. the second subtraction is at
$\mu^2=Q^2$:
$${\cal F}_{(-)}({\mu^2\over Q^2})-{\cal F}_{(-)}(0) = a_1{\mu^2\over Q^2} + \int_{0}^\infty 
{dk^2\over k^2}\ \left[{-{\mu^2\over k^2}\over1+{\mu^2\over k^2}}+ {\mu^2 k^2\over
(k^2+Q^2)^2}\right]\
\varphi_{(-)}\left(k^2\over Q^2\right)\eqno(B.9)$$ where:
$$a_1\equiv{\cal F}_{(-)}'(1)$$ one finds:
$$R_{reg,<}^{PT}(Q^2)=-a_1 I_1 {\Lambda^2\over Q^2} + \int_{0}^\infty{dk^2\over k^2}\
\left[\alpha_{reg,<}^{PT}(k^2)-I_1 {\Lambda^2 k^2\over (k^2+Q^2)^2}\right]\
\varphi_{(-)}\left(k^2\over Q^2\right)\eqno(B.10)$$ and:
$$R_<^{PT}(Q^2)=-a_1 I_1^{PT} {\Lambda^2\over Q^2} + \int_{0}^\infty{dk^2\over k^2}\
\left[\alpha_<^{PT}(k^2)-I_1^{PT} {\Lambda^2 k^2\over (k^2+Q^2)^2}\right]\
\varphi_{(-)}\left(k^2\over Q^2\right)\eqno(B.11)$$ Hence:
$$\delta R_{PT}(Q^2)=-a_1 b_1^{PT} {\Lambda^2\over Q^2} + \int_{0}^\infty{dk^2\over k^2}\
\left[\delta\alpha_s^{PT}(k^2)-b_1^{PT} {\Lambda^2 k^2\over (k^2+Q^2)^2}\right]\
\varphi_{(-)}\left(k^2\over Q^2\right)\eqno(B.12)$$ The subtraction term in the integrand is
easily managed, since eq.(B.12) can be put in the form:
\begin{eqnarray*}\delta R_{PT}(Q^2)&=&-B_1 b_1^{PT} {\Lambda^2\over Q^2} +
\int_{0}^{Q^2}{dk^2\over k^2}\ \delta\alpha_s^{PT}(k^2)\ \varphi_{(-)}\left(k^2\over Q^2\right) \\
& &+\int_{Q^2}^\infty{dk^2\over k^2}\ \left[\delta\alpha_s^{PT}(k^2)-b_1^{PT} {\Lambda^2\over
k^2}\right]\ \varphi_{(-)}\left(k^2\over Q^2\right)\ \ \ \ \
\ \ \ \ \ \ \ \ \ \ \ \ \ \ \ \ \ \ \ \ \ (B.13)\end{eqnarray*} where:
$$B_1=a_1 + \int_{0}^{Q^2}{dk^2\over k^2}\ {k^2 Q^2\over (k^2+Q^2)^2}\
\varphi_{(-)}\left(k^2\over Q^2\right) + \int_{Q^2}^\infty{dk^2
\over k^2}\ \left[{k^2 Q^2\over (k^2+Q^2)^2} - {Q^2\over k^2}\right]\ \varphi_{(-)}\left(k^2\over
Q^2\right)$$ is a number.

In the one-loop coupling case,  eq.(B.4), (B.8) and (B.12) reproduce again the Beneke-Braun
result eq.(6.29),   provided one  understands
${\cal F}$ as being the analytic continuation of ${\cal F}_{(-)}$ to the $\mu^2/Q^2<~0$ region:
$$\delta R_{PT}(Q^2)=-{1\over\beta_0} \left[{\cal F}_{(-)}\left(-\ {\Lambda^2\over
Q^2}\right)-{\cal F}_{(-)}(0)\right]\eqno(B.14)$$ Eq.(B.14) follows from the observation (which
parallels a similar one in section 6) that the  dispersion relations for
${\cal F}_{(-)}({\mu^2\over Q^2})-{\cal F}_{(-)}(0)$ (eq.(B.1), (B.5) and (B.9)) may be seen as
peculiar cases of the formulas for $\delta R_{PT}(Q^2)$ (eq.(B.4), (B.8) and (B.12) respectively)
, with the substitutions: $\Lambda^2
\rightarrow -\mu^2 $ and 
$\delta\alpha_s^{PT}(k^2)\rightarrow {-{\mu^2\over k^2}\over1+{\mu^2\over k^2}}$. Note also the
results of this section  are easily extended to the rather general case where ${\cal
F}_{(-)}(\mu^2/Q^2)$ can be written as the sum of a function which satisfies a dispersion
relation (hence has no complex singularities) and a function analytic around the origin - a
generalized ``subtraction term'' (but which may have complex singularities at finite distance
from the origin).\\

\noindent{\bf B2 \ \ \ \ \   Non perturbative power corrections}\\

\noindent Formulas which allow an explicit connection of $\delta R_{NP}$ with 
$\delta\alpha_s^{NP}$  also exist, similar to those of Appendix B1 for $\delta R_{PT}$ (but valid
only at large $Q^2$). Indeed, starting from the general expression:
$$\delta R_{NP}(Q^2) = \int_0^{Q^2}{d\mu^2\over\mu^2}\ \delta\alpha_{eff}^{NP}(\mu^2)\ \dot{{\cal
F}}_{(-)}({\mu^2\over Q^2})+\int_{Q^2}^\infty{d\mu^2\over\mu^2}\ \delta\alpha_{eff}^{NP}(\mu^2)\
\dot{{\cal F}}_{(+)}({\mu^2\over Q^2})\eqno(B.15)$$ the assumed exponential decrease of
$\delta\alpha_{eff}^{NP}(\mu^2)$ allows to write, up to exponentially small corrections at large
$Q^2$  :
\begin{eqnarray*}\ \ \ \ \ \ \ \ \
\delta R_{NP}(Q^2)&\simeq&\int_0^{Q^2}{d\mu^2\over\mu^2}\ \delta\alpha_{eff}^{NP} (\mu^2)\
\dot{{\cal F}}_{(-)}({\mu^2\over Q^2})\\ &\simeq&\int_0^\infty{d\mu^2\over\mu^2}\
\delta\alpha_{eff}^{NP} (\mu^2)\ \dot{{\cal F}}_{(-)}({\mu^2\over Q^2})\ \ \ \ \
\ \ \ \ \ \ \ \ \ \ \ \ \ \ \ \ \ \ \ \ \ \ \ \ \ \ \ \ \ \ \ (B.16)\end{eqnarray*} Assume now
${\cal F}_{(-)}({\mu^2\over Q^2})$ satisfies the twice subtracted dispersion relation at zero
$\mu^2$ eq.(B.5). Then one finds after substitution in eq.(B.16), and using eq.(2.9):
$$\delta R_{NP}(Q^2)\simeq -a_0 b_1^{NP} {\Lambda^2\over Q^2} + \int_{0}^\infty{dk^2\over k^2}\
\left[\delta\alpha_s^{NP}(k^2)-b_1^{NP} {\Lambda^2\over k^2}\right]\ \varphi_{(-)}\left(k^2\over
Q^2\right)\eqno(B.17)$$ Eq.(B.17) has {\em exactly} the same form as eq.(B.8) for $\delta
R_{PT}(Q^2)$, with the substitutions $\delta\alpha_s^{PT}\rightarrow
\delta\alpha_s^{NP}$, and
$b_1^{PT}\rightarrow b_1^{NP}$! A result similar to eq.(B.12), with the same substitutions, also
holds if one starts from the dispersion relation eq.(B.9) with a subtraction  at non-zero
$\mu^2$, and of course the analogue of eq.(B.4), i.e.:
$$\delta R_{NP}(Q^2)\simeq \int_{0}^\infty{dk^2\over k^2}\ \delta\alpha_s^{NP}(k^2)\
\varphi_{(-)}\left(k^2\over Q^2\right)\eqno(B.18)$$  holds if only one subtraction at
$\mu^2=0$ is necessary (eq.(B.1)). It follows that similar formulas are also valid for the {\em
total} power correction
$\delta R=\delta R_{PT}+\delta R_{NP}$, with the substitutions  
$\delta\alpha_s^{PT}\rightarrow\delta\alpha_s=\delta\alpha_s^{PT}+\delta\alpha_s^{NP}$ and 
$b_1^{PT}\rightarrow b_1=b_1^{PT}+ b_1^{NP}$.

\noindent If one now requires $\delta\alpha_s^{NP}$ (resp. $\delta\alpha_s$) be restricted to low
$k^2$, i.e. that $b_n^{NP}=0$ (resp. $b_n=0$),  then the subtraction terms in eq.(B.17)  vanish,
since they are proportionnal to $b_1^{NP}$ (resp. $b_1$)  and one ends up with eq.(B.18) (or a
similar result for $\delta R$) whatever the subtractions needed for ${\cal F}_{(-)}$. Comparing
eq.(B.16) and (B.18), one gets the identity (valid only if $\delta\alpha_s^{NP}$ is restricted to
low $k^2$):
$$\int_{0}^\infty{dk^2\over k^2}\ \delta\alpha_s^{NP}(k^2)\ \varphi_{(-)}\left(k^2\over
Q^2\right)=\int_0^\infty{d\mu^2\over
\mu^2}\ \delta\alpha_{eff}^{NP} (\mu^2)\ \dot{{\cal F}}_{(-)}({\mu^2\over Q^2})\eqno(B.19)$$ from
which eq.(7.36) and (7.41) may also be derived.

\newpage
\noindent{\bf References}\\

\noindent 1. G.P. Korchemsky and G. Sterman, in {\sl Proc. 30th Rencontres de Moriond,
Meribel-les-Allues, France, 1995} [hep-ph/9505391]; R. Akhoury and V.I. Zakharov, in {\sl
QCD'96}, [hep-ph/9610492]; V.M. Braun, NORDITA-96-65P [hep-ph/9610212]; M. Beneke,  SLAC-PUB-7277
[hep-ph/9609215]; P. Nason and M.H. Seymour, {\sl Nucl. Phys.} {\bf B454} (1995) 291.\\

\noindent 2. B.R. Webber, {\sl Phys.Lett.} {\bf B339} (1994) 148.\\

\noindent 3. I.I. Bigi, M.A. Shifman, N.G. Uraltsev and A.I. Vainshtein, {\sl Phys.Rev.}{\bf D50}
(1994) 2234.\\

\noindent 4. M. Beneke, V.M. Braun and V.I. Zakharov, {\sl Phys.Rev.Lett.} {\bf 73} (1994) 3058.\\

\noindent 5. M. Beneke and V.M. Braun, {\sl Nucl. Phys.} {\bf B454} (1995) 253.\\

\noindent 6. M. Beneke and V.M. Braun, {\sl Phys.Lett.} {\bf B348} (1995) 513; P. Ball, M. Beneke
and V.M. Braun, {\sl Nucl. Phys.} {\bf B452} (1995) 563.\\

\noindent 7. Yu.L. Dokshitzer, G. Marchesini and B.R. Webber,  {\sl Nucl. Phys.} {\bf B469} 
(1996) 93.\\

\noindent 8. Yu.L. Dokshitzer and B.R. Webber,  {\sl Phys.Lett.} {\bf B352} (1995) 451.\\

\noindent 9. Yu.L. Dokshitzer, V.A. Khoze and S.I. Troyan, {\sl Phys.Rev.} {\bf D53} (1996) 89.\\

\noindent 10. P. Redmond, {\sl Phys.Rev.}{\bf 112} (1958) 1404; P. Redmond and J.L. Uretsky, {\sl
Phys.Rev.Lett.} {\bf 1} (1958) 147.\\

\noindent 11. N.N. Bogoliubov, A.A. Logunov and D.V. Shirkov, {\sl Sov. Phys. JETP} {\bf 37}(10)
(1960) 574.\\

\noindent 12. D.V. Shirkov and I.L. Solovtsov, {\sl JINR Rapid Comm.}, No.2[76]-96, 5
[hep-ph/9604363]; hep-ph/9704333.\\

\noindent 13. K.A. Milton and I.L. Solovtsov, hep-ph/9611438.\\ 

\noindent 14. Yu.L. Dokshitzer and D.V. Shirkov, {\sl Zeit. Phys.} {\bf C67} (1995) 449.\\

\noindent 15. S.J. Brodsky, G.P. Lepage and P.B. Mackenzie, {\sl Phys.Rev.} {\bf D28} (1983)
228.\\

\noindent 16. H.J. Lu and C.A.R. Sa de Melo, {\sl Phys.Lett.} {\bf B273} (1991) 260 ( Erratum
ibid.{\bf B285} (1992) 399 ); H.J.Lu, {\sl Phys.Rev.} {\bf D45} (1992) 1217; ibid SLAC-0406
(Ph.D. Thesis) (1992).\\

\noindent 17. D.J. Broadhurst and A.G. Grozin, {\sl Phys.Rev.} {\bf D52} (1995) 4082.\\

\noindent 18. M. Neubert, {\sl Phys.Rev.} {\bf D51} (1995) 5924; CERN-TH.7524/94
[hep-ph/9502264].\\

\noindent 19. N.J. Watson, CPT-96-P-3347 [hep-ph/9606381].\\

\noindent 20. M.A. Shifman, A.I. Vainshtein and V.I. Zakharov, {\sl Nucl. Phys.} {\bf B147}  
(1979) 385.\\

\noindent 21. A.H. Mueller, {\sl Phys.Lett.} {\bf B308} (1993) 355; in {\sl QCD-20 Years Later},
vol.1 (World Scientific, Singapore,1993).\\

\noindent 22. G. Grunberg, {\sl Phys.Lett.} {\bf B325} (1994) 441; {\bf B349} (1995) 469.\\

\noindent 23. G. Grunberg,  {\sl Phys.Lett.} {\bf B372} (1996) 121; hep-ph/9608375.\\

\noindent 24. Yu.L. Dokshitzer and N.G. Uraltsev, {\sl Phys.Lett.} {\bf B380}  (1996) 141.\\

\noindent 25. G. Grunberg,  {\sl Phys.Lett.} {\bf B304} (1993) 183; in {\sl Quantum Field
Theoretic Aspects of High Energy Physics, Kyffhauser,Germany,september 1993}.\\

\noindent 26. M. Beneke, {\sl Nucl.Phys.} {\bf B405} (1993) 424.\\

\noindent 27. L.S. Brown and L.G. Yaffe, {\sl Phys.Rev.} {\bf D45} (1992) 398;  L.S. Brown, L.G.
Yaffe and C. Zhai, {\sl Phys.Rev.} {\bf D46} (1992) 4712.\\

\noindent 28. G. Grunberg, {\sl Phys.Rev.} {\bf D46} (1992) 2228.\\

\noindent 29. N.G. Uraltsev, private communication.\\

\noindent 30. A.H. Mueller, {\sl Nucl.Phys.} {\bf B250}  (1985) 327.\\

\noindent 31. V.A. Novikov et al., {\sl Phys.Rep.} {\bf 116} (1984) 104; {\sl Nucl. Phys.} {\bf
B249} (1985) 445.\\

\noindent 32. V.I. Zakharov, {\sl Nucl.Phys.} {\bf B385} (1992) 452.\\

\noindent 33. See e.g. D.V. Widder, {\bf The Laplace Transform}, Princeton University Press
(1946).\\

\noindent 34. A.I Alekseev and B.A. Arbuzov, hep-ph/9704228.\\

\noindent 35. Yu.L. Dokshitzer and B.R. Webber, Cavendish-HEP-97/2 [hep-ph/9704298] and
references therein.\\

\noindent 36. G. Altarelli, P. Nason and G. Ridolfi, {\sl Zeit.Phys.} {\bf C68} (1995) 257.\\

\noindent 37. O.P. Solovtsova, {\sl JETP Lett.} {\bf 64},No.10 (1996) 714.\\

\noindent 38. A. Hocker, in {\sl Fourth International Workshop on Tau Lepton Physics, Estes Park,
Colorado, USA, September 1996},  LAL 96-95 [hep-ex/9703004].\\

\noindent 39. F. di Renzo, E. Onofri, G. Marchesini, {\sl Nucl. Phys.} {\bf B457} (1995) 202.\\

\noindent 40. G. Marchesini, private communication.\\

\noindent 41. F.David, {\sl Nucl. Phys.} {\bf B234} (1984) 237; {\bf B263} (1986) 637.\\

\noindent 42. G. Martinelli and C.T. Sachrajda, {\sl Nucl. Phys.} {\bf B478} (1996) 660.\\

\noindent 43. M. Beneke and V.I Zakharov, {\sl Phys.Rev.Lett.} {\bf 69} (1992) 2472.\\

\noindent 44. G. Altarelli, in {\sl Proc. Third Workshop on Tau Lepton Physics} (Montreux,1994).\\

\noindent 45. E.B. Bogomolny and V.A. Fateyev, {\sl Phys.Lett.} {\bf B71} (1977) 93.

\end{document}